\documentclass[]{aa}

\usepackage[normalem]{ulem}
\usepackage{graphicx}
\usepackage{txfonts}
\usepackage{natbib}
\bibpunct{(}{)}{;}{a}{}{,}
\usepackage{scrextend}			
\usepackage{tikz}
\usepackage{subcaption}
\usepackage{multirow}
\usepackage{hyperref}
\usepackage{amsmath}  
\usepackage{xfrac}   
\usepackage{amssymb}	
\usepackage{graphicx}
\usepackage{amsmath}  
\usepackage{xfrac}   
\usepackage{amssymb}	
\usepackage{ dsfont }
\usepackage{txfonts}
\usepackage{mwe,tikz}\usepackage[percent]{overpic}

\begin{document}

  \title{A unified accretion-ejection paradigm for black hole X-ray binaries}

  \subtitle{II. Observational signatures of jet-emitting disks}

  \author{G. Marcel\inst{1}
          \and
          J. Ferreira\inst{1} 
          \and 
          P-O. Petrucci\inst{1}
          \and
          G. Henri\inst{1}
          \and
          R. Belmont\inst{2} \fnmsep \inst{3}
          \and
          M. Clavel\inst{1,4}
          \and
          J. Malzac\inst{2} \fnmsep \inst{3}
          \and
          M. Coriat\inst{2} \fnmsep \inst{3}
          \and
          S. Corbel\inst{5}
          \and
          J. Rodriguez\inst{5}
          \and
          A. Loh\inst{5}
          \and
          S. Chakravorty\inst{6}
          \and
          S. Drappeau\inst{2} \fnmsep \inst{3}
          }
         
   \institute{Univ. Grenoble Alpes, CNRS, IPAG, 38000 Grenoble, France\\
              \email{gregoire.marcel@univ-grenoble-alpes.fr or gregoiremarcel26@gmail.com}
              \and
              Universite de Toulouse; UPS-OMP; IRAP; Toulouse, France
              \and
              CNRS; IRAP; 9 Av. colonel Roche, BP 44346, F-31028 Toulouse cedex 4, France
              \and
              Space Sciences Laboratory, 7 Gauss Way, University of California, Berkeley, CA 94720-7450, USA
              \and
              Laboratoire AIM (CEA/IRFU - CNRS/INSU - Universite Paris Diderot), CEA DSM/IRFU/SAp, F-91191 Gif-sur-Yvette, France
              \and 
              Department of Physics, Indian Institute of Science, Bangalore 560012, India
             }

   \date{Received October 10, 2017; accepted March 8, 2018}

   \abstract
   {X-ray binaries display cycles of strong activity during which their luminosity varies across several orders of magnitude. The rising phase is characterized by a hard X-ray spectrum and radio emission due to jets (hard state), whereas the declining phase displays a soft X-ray spectrum and no jet signature (soft state). The origin of these correlated accretion-ejection and spectral hysteresis cycles is still under investigation.}       
   {We elaborate on the paradigm proposed in \citet{Ferreira06}, where the increase and decrease in the disk accretion rate is accompanied by a modification of the disk magnetization $\mu$, which in turn determines the dominant torque allowing accretion. For $\mu$ greater than some threshold, the accretion flow produces jets that vertically carry away the disk angular momentum (jet-emitting disk, or JED mode), whereas for smaller $\mu$, the turbulence transfers the disk angular momentum outward in the radial direction (standard accretion disk, or SAD mode). The goal of this paper is to investigate the spectral signatures of the JED configurations.} 
   {We have developped a two-temperature plasma code that computes the disk local thermal equilibria, taking into account the advection of energy in an iterative way. Our code addresses optically thin/thick transitions, both radiation and gas supported regimes, and computes in a consistent way the emitted spectrum from a steady-state disk. The optically thin emission is obtained using the \textsc{BELM} code, which provides accurate spectra for bremsstrahlung and synchrotron emission processes as well as for their local Comptonization.}
   {For a range in radius and accretion rates, JEDs exhibit three thermal equilibria, one thermally unstable and two stable: a cold (optically thick and geometrically thin) and a hot (optically thin and geometrically thick) equilibrium. From the two thermally stable solutions, a hysteresis cycle is naturally obtained. However, standard outbursting X-ray binary cycles cannot be reproduced. Another striking feature of JEDs is their ability to reproduce luminous hard states. At high accretion rates, JEDs become slim, where the main cooling is advection.}
{When the loss of angular momentum and power in jets is consistently taken into account (JED mode), accretion disks have spectral signatures that are consistent with hard states, up to high luminosities. When no jet is present (SAD mode), the spectral signature is consistent with the soft state. These two canonical spectral states of black hole binaries can be explained in terms of two completely different dynamical solutions, namely JED and SAD. The observed spectral cycles can therefore be directly understood in terms of dynamical transitions from one accretion mode to another.  These transitions must involve states where some regions emit jets and others do not, however, which argues for hybrid disk configurations.}


   \maketitle
%

\section{Introduction}


A huge amount of data at all wavelengths has been collected in the past 20 years on black hole X-ray binaries, hereafter XrBs \citep[for a global review see][]{Dunn10}. These objects spend most of their time in quiescence at very low accretion rates, but occasionally, they produce outbursts that last from a few months to a year. Their flux then rises by several orders of magnitude across the whole electromagnetic spectrum \citep[see, e.g.,][for recent reviews]{Corbel04, Fender06, Remillard06, Done07}. During an outburst, XrBs show very different spectral and temporal states that can be easily distinguished in a hardness-intensity diagram (HID) where the X-ray luminosity is plotted versus the hardness ratio of the X-ray spectrum \citep[see, e.g.,][]{Homan01, Fender04}. The evolutionary track produces a typical q-shaped figure that reveals a hysteresis: outbursting XrBs have two distinct spectra with the same X-ray luminosity above $1-2\%$ Eddington luminosity. At the beginning of the outburst, the system is in the so-called hard state: the spectrum has a hard power-law shape up to a few tens to hundreds of keV, requiring a very hot, optically thin plasma (referred to as the "corona"). Then, when the system reaches high luminosities (up to a few tens of the Eddington luminosity), it transits within a few days through a bright intermediate state into the so-called soft state (referred to as the "cold disk"). In this state, the spectrum is dominated by strong and soft X-ray emission, which is commonly interpreted as thermal emission from an optically thick geometrically thin accretion flow. In the latter state, the luminosity starts to decrease and the system returns to the hard state, transiting through a faint intermediate state. The luminosities at which a system transits from hard to soft states are several times higher than the luminosity of the reverse transition \citep[see Appendix in][]{Dunn10}.

Similarly to some active galactic nuclei, XrBs also show evidence of jets \citep[e.g.,][]{Mirabel92, Fender97} that is usually observed in the radio band but that can, at least in extreme cases, contribute significantly to the X-ray emission \citep{Corbel02, Rodriguez08}. The flat-to-inverted slope detected in radio and IR fits the spectrum expected from a stratified self-absorbed compact synchrotron jet very well \citep[see][]{BK79}. These jets are systematically observed in the hard states, showing a tight correlation between the radio and the X-ray emission, which is attributed to the jet and the accretion flow, respectively \citep{CorbelFender02, Corbel00, Corbel03, Corbel13, Gallo03, Merloni03, Falcke04}. When the system transits to the soft state, the radio/IR emission is strongly reduced until it becomes undetectable \citep[e.g.,][]{Coriat09, Coriat11}. This suggests the disappearance of the jet component when the accretion flow becomes geometrically thin and optically thick. At the end of the outburst, when the system returns to the hard state, the radio/IR emission that
is due to the jet reemerges. Clearly, there is a tight correlation between the accretion flow spectral states and the emission properties of the jets, which argues for a direct dynamical link between these two.\\  

The first attempts to address this complex behavior were mostly focused on the thermal properties of the accretion
flow. In the scenario proposed by \citet{Esin97}, the inner accretion flow would be in a radiatively inefficient, optically thin thermal state during the low-luminosity hard state and would progressively change into a radiatively efficient, optically thick state in the luminous soft state. In terms of dynamics, the accretion flow was assumed to transit from an inner advection-dominated accretion flow or ADAF \citep{Ichimaru77, Rees82, Narayan94} to an outer SAD \citep[][]{SS73}. This was the first proposition to provide an interesting explanation for the surprising behavior of XRBs. However, this scenario faced open questions. One of them was explaining the observed hysteresis cycles. Why would XrBs follow a different thermal path while declining? A second question was related to the very existence of ADAF solutions, in particular at high luminosities, as required by observations. A third question was the failure to explain the self-confined jets during the hard states. We examine these difficulties in more detail.

The scenario proposed by \citet{Esin97} triggered many theoretical studies on how the accretion flow might transit outside-in from the thin accretion disk solution to the ADAF-like solution. 
One possibility was the "strong ADAF principle", formulated by \citet{NarayanYi95}, which states that the accretion disk would always "choose" the optically thin ADAF solution when several thermal solutions are possible. Although the strong ADAF principle lacks physical grounds, this idea nevertheless allowed
interesting studies of the physics at the transition radius $R_t$ between these two solutions \citep[e.g.,][]{Honma96, Abramowicz98, Kato00, Gracia03}. An explanation for this transition was then introduced by \citet{Meyer00} and \citet{Rozanska00}, based on the idea of disk evaporation through a coronal flow \citep[see, e.g.,][]{Meyer94}. The transition radius $R_t$ would naturally arise when the mass loss due to this evaporation mechanism matches the disk accretion rate \citep[see also][]{Spruit02}. Later, taking into account the illumination effect that photons from the central light source have on the evaporation rate of the outer accretion flow, \citet{Meyer05} provided a fairly convincing model for the hysteresis cycle \citep[see also][and references therein]{Meyer14}. However, this scenario relies on the dynamical ADAF solution, which raises several problems.

Any ADAF solution requires two conditions: (1) a low-density plasma, so that the disk is not thermalized, and (2) a negligible fraction $\delta$ of the released magnetohydrodynamic (MHD) turbulent energy that heats the electrons. 
The first condition is never fulfilled in the high-luminosity hard states \citep[see, e.g.,][]{Oda12}. Nevertheless, \citet{Yuan01} showed that optically thin, advection-dominated flows could still survive at a much higher luminosity than ADAFs; such flows where termed hot luminous flows or LHAF (for luminous hot accretion flow) and were initially obtained assuming $\delta=0$. The existence of such solutions lies in the fact that the advection of internal energy plays the role of a local heating instead of a cooling: the inner accretion flow is thinner and radiates more strongly. The change of sign of the advection term (from cooling to heating) occurs only in the inner regions in this global solution subject to three constraints: the no-torque, sonic regularity, and an outer boundary condition. We refer to \citet{Yuan00} for further details.
The second condition regarding ADAFs, an electron heating parameter $\delta <<1$, implies that only the ions are directly heated and transfer their energy through Coulomb collisions to the electrons. Much work has been done on the value of $\delta,$ and it appears to be highly dependent on the magnetic field strength and on
how the MHD turbulent energy is dissipated. More precisely, it depends on the nature of the dominant resonant turbulent waves and whether reconnection is present \citep[see, e.g.,][]{Gruzinov98, Quataert99, Bisnovatyi00, Quataert02, Lehe09}. The currently accepted value is $\delta \sim 0.1-0.5$ \citep[][]{Yuan14} and closer to 0.5 for near equipartition fields, which would rule out ADAF or ADIOS models \citep[see discussions in][]{Yuan14, BB99}.

For LHAFs, \citet{Xie12} revisited the model by assuming $\delta\sim 0.1-0.5$ but also including a mass loss $\dot M(R) \propto R^s$ with $s=0.4$ due to a massive outflow. Such a high value of $s$ leads to a significant modification of the accretion flow density profile, which drastically decreases the efficiency of the compressional work on ions. Because LHAFs mostly rely on this advection heating, the parameter regime for their existence is actually reduced. The authors found that beyond a few percent Eddington luminosity, only the optically thick, geometrically thin accretion flow solution is available. Again, although the model reproduces many aspects of X-ray binary behavior, high-luminosity hard states seem beyond
reach. \\ %
 
Yet another critical ingredient has been neglected so far: jet production while in hard states and quenching in soft states. None of the models mentioned above includes jet formation within their disk dynamical description, which completely decorrelates the accretion flow and the jet properties. At best, mass loss is taken into account, but the dominant torque remains the torque due to turbulence, and all released energy is either advected or released as radiation.

The reason generally invoked is that jets are assumed to arise only from the black hole ergosphere, following the seminal paper of \citet{BZ77}, hereafter BZ. In this scenario, BZ jet power arises from the black hole rotational power. The surrounding disk serves only as a mass reservoir and as an electric conductor, maintaining toroidal electric currents and possibly fueling the black hole magnetosphere with some magnetic flux \citep[see, e.g.,][and references therein]{Tchekhovskoy11, McKinney12, Lasota14}. However, these models do not explain the hysteresis cycles observed in XrBs. One possibility would be that the magnetic field required to launch BZ jets can only be maintained when the accretion flow is geometrically thick (hard state), whereas it would diffuse away when it transits to the optically thick (soft state) and geometrically thin regime \citep[e.g.,][]{Igumenshchev09, Penna10, Sikora13, Piran15}. However, some work remains to be done to compare this idea to observations. In addition, BZ jets require a rotating black hole. The task remains to reconcile such a scenario with similar accretion-ejection hysteresis cycles observed from neutron stars or even white dwarfs \citep{Kording07, Migliari07, MillerJones10, Kording14, MunozDarias14}.

Another possibility would be that the magnetic field required to launch jets is generated by a disk dynamo \citep[e.g.,][]{Meier01, Livio03, King04, BegelmanArmitage14, Begelman15, Salvesen16, Riols17}. However, jet production requires a strong vertical field, and to date, no $B_z$ field amplification has ever been observed in 3D global simulations of accretion flows. Jets are never observed unless some initial vertical (large-scale) magnetic field is initially included in the simulation \citep{Beckwith08}.  A different dynamo mechanism has been proposed by \citet{Contopoulos98}. In this cosmic battery, the source of the magnetic field is the azimuthal electric current associated with the Poynting-Robertson drag on the electrons of the accreting plasma. This promising mechanism deserves further development so as to be compared to observations. The model relies on a transition radius between an inner optically thin and an outer optically thick accretion flow \citep{Contopoulos15}. The coupling between the dynamo action, jet launching, and the thermal properties of the accretion flow remain to be assessed. \\

The other way to produce self-confined jets is to tap the mechanical energy released in the accretion flow itself, following the seminal paper of \citet{BP82}, hereafter BP. These BP jets also require a large-scale $B_z$ magnetic field threading the disk. The first studies concentrated on cold flows, purely magnetically driven jets, that carry away mass, energy, and angular momentum belonging to the disk. As a consequence, they deeply affect the underlying disk structure \citep{Ferreira93a, Ferreira95}, defining a new class of accretion flows. In this new accretion mode solution, termed jet-emitting disks (hereafter JEDs), the dominant torque is induced by the jets. In order to produce super-Alfv\'enic cold BP jets, smaller than but near to equipartition fields are found to be necessary \citep{Ferreira97, Casse00a}. The origin of such a strong magnetic field over a large radial extent within the disk is assumed to be a consequence of the complex interplay between field advection and diffusion. If the $B_z$ magnetic field is strong enough, a JED can be established, whereas if the magnetic field is smaller than a given threshold value, then no jet launching is assumed to be possible and the accretion takes place through MHD turbulence, as in the SAD mode.  

Based on these physical ingredients, we proposed a paradigm for the hysteresis cycles of X-ray binaries in \citet{Ferreira06}, hereafter paper I. This paradigm assumes that the disk is made of radial extended zones where one of these two accretion modes, JED or SAD, is established. It has been argued that the transition between these two modes depends mostly on the disk magnetization \citep[paper I,][]{Petrucci08}. We thus propose to view the inner accretion flow (and its spectral signature) as a dynamical system that responds almost instantaneously to the evolution of two independent control parameters: the disk accretion rate, and the disk magnetization. A given parameter set therefore defines a disk configuration \citep[][]{Petrucci08, Petrucci10}. The aim of the present study is to compute the spectral signatures associated with a JED configuration as
accurately as possible, and a companion paper will aim at developing the hybrid JED-SAD configuration.\\

The paper is organized as follows. Section 2 describes how the thermal disk balance is computed, using a new two-temperature plasma code that also provides the spectral energy distribution
of the disk. This is a generalization of the work by \citet{Petrucci10}, which was done solely for a one-temperature plasma. Section 3 describes the three classes of thermal states that have been found and their associated spectra. The influence of the JED dynamical parameters is extensively studied in Section 4.  The next two sections focus on two striking properties of JED configurations. Section 5 investigates the characteristic JED thermal hysteresis cycles and compares them with the observed typical q-shaped cycles. Section 6 studies the most favorable dynamical JED parameters in greater detail that allow reproducing hard states up to high luminosities. Section 7 discusses several caveats of our analysis and summarizes our findings.

\section{Thermal structure of accretion disks}
\label{sec:Structure}

\subsection{Dynamical disk configurations}

We consider an axisymmetric (cylindrical coordinates) accretion disk around a black hole of mass $M$. Throughout the paper, the calculations are made within the Newtonian approximation. We define $R$ the radius, $H(R)$ the half-height of the disk, $\varepsilon(R) = H/R$ its aspect ratio, $u_R$ the radial (accretion) velocity and $\Sigma = \rho_0 H$ the vertical column density with $\rho_0$ the mid-plane density. Moreover, and for the sake of simplicity, the disk is assumed to be always Keplerian, with a local angular velocity $\Omega \simeq \Omega_K = \sqrt[]{G M R^{-3}}$, where $G$ is the gravitational constant. These approximations clearly
have an impact, but addressing these points is beyond the scope of the present paper. 

For simplicity, the disk is assumed to be in global steady-state so that any variation of the disk accretion rate $\dot{M}(R) =  - 4 \pi R u_R \Sigma$ would be due to mass loss (jets or winds) from the disk alone, see eq.~(\ref{eq:defXi}) below.

The disk is assumed to be thread by a large-scale vertical magnetic field $B_z(R)$. As for any other disk quantity, the local magnetic field is assumed to be stationary on dynamical timescales (Keplerian orbital time), an evolution remaining possible on longer (accretion) timescales.

A JED configuration is described by the following elements:\\
(1) The central object has a black hole mass $M$ and an innermost radius $R_{in}$ (a proxy for the black hole spin, which is not considered here),\\
(2) the disk accretion rate feeding the black hole $\dot M_{in}= \dot{M}(R_{in})$ from the innermost radius, and\\
(3) the radial extent where the JED is established: between $R_{in}$ and the outer disk radius  $R_{out}$.\\
 
In the following, we adopt the dimensionless scalings $r = R/R_g$, $h = H/R_g= \varepsilon r$, where $R_g = GM/c^2$, $m = M/M_{\odot}$, and $\dot{m} = \dot{M}/\dot{M}_{Edd}$ and $\dot{M}_{Edd} = L_{Edd} / c^2$ is the Eddington accretion assuming maximum efficiency and $L_{Edd}$ is the Eddington luminosity. With these scalings, we explore the following range of values for XrB accretion disk physical parameters:
\begin{itemize}
        \item Disk internal radius, $1 \leq r_{in} \leq 6$
        \item Disk external radius, $r_{in} < r_{out} \leq 10^6$
        \item Disk accretion rate, $10^{-6} \leq \dot{m}_{in} \leq 10^2$
\end{itemize}

\subsection{Model parameters}

The obvious dynamical difference provided by the JED solutions (compared with the SAD solutions) is jet production. This is a very important difference since accretion and ejection are inter-dependent.
In a JED mode, the disk launches self-confined jets and the disk accretion rate writes
\begin{equation}
\dot{m}\left(r\right) = \dot m_{in} \left ( \frac{r}{r_{in}} \right )^\xi \label{eq:defXi}
,\end{equation}
\noindent where $\dot m_{in}= \dot M_{in} c^2/L_{Edd}$ is the normalized disk accretion rate feeding the central black hole and $\xi$ is the disk ejection efficiency \citep[see][]{Ferreira93a}. In contrast to all alternative magnetically driven disk wind models, the value of the ejection efficiency $\xi$ was computed as function of the magnetic field strength and depending on whether thermal effects are relevant at the base of the outflow. For cold tenuous fast jets, a typical value is $\xi \sim 0.01$ \citep[see][]{Ferreira97}, whereas both denser and slower warm (magnetothermal) winds are obtained with $\xi > 0.1$ and up to 0.5 \citep[see][]{Casse00b}. Since the disk luminosity varies by many orders of magnitude during an outburst, it is natural to use $\dot m_{in}$ as a control parameter varying in time (on long timescales). 

The global energy budget of a quasi-Keplerian accretion disk established between two radii $r_1$ and $r_2>r_1$ can be written as  
\begin{equation} 
P_{acc} = 2 P_{jet} + P_{adv} + P_{disk}
,\end{equation}
where $P_{acc}$ is the mechanical power released by accretion from $r_2$ to $r_1$, $P_{jet}$ is the MHD power feeding each jet (factor $2$) produced by each side of the disk in the range of radii, $P_{adv}$ the thermal power conveyed by the accretion flow from $r_2$ to $r_1$ , and $P_{disk}$ the total disk luminosity. In this energy budget, the total power brought into the disk by turbulence is assumed to be negligible with respect to the released mechanical power (no-torque condition at the innermost radius). We define the jet power fraction as
\begin{equation}
 b = \frac{2 P_{jet}}{P_{acc}} \label{eq:defB}
\end{equation}
since the JED transfers a significant fraction of the accretion energy into the jets because in a JED mode, jets carry the entire
disk angular momentum away and diminish the fraction of energy to be radiated. JEDs therefore have a lower radiative efficiency than SADs.

The disk magnetization is locally defined as
\begin{equation}
\mu = \frac{B_z^2/\mu_0}{P_{tot}} = \frac{B_z^2 / \mu_0}{P_{gas} + P_{rad}}
\end{equation}
\noindent and measured at the disk mid-plane. Here, $B_z$ is the vertical magnetic field, $\mu_0$ the vacuum permeability,
and $P_{gas}$ and $P_{rad}$ are the plasma kinetic and radiation pressure, respectively. We note that in the mean field approach used by \citet{Ferreira97} and related works, the contribution of turbulent magnetic fields has been neglected to compute accretion-ejection flows. Numerical simulations of MRI do show the development of a turbulent pressure \citep[see, e.g.,][and references therein]{Salvesen16}, but these simulations were made with mostly small magnetic fields (although one simulation was made with $\mu= 0.2$) and, more importantly, in a shearing box so that no BP jets could be launched (by construction). It is therefore not clear whether such a turbulent pressure would remain at this level in JEDs, which require near equipartition magnetic fields. In any case, this might be a possible limitation of the current model and should be kept in mind. 

The value of the accretion speed $u_R$ depends on the dominant torque acting upon the disk material. Its strength can be measured in the plane of the disk using the sonic Mach number \citep[see][]{Casse00a}
\begin{equation}
m_s = \frac{-u_R}{c_s} = \frac{-u_R}{\Omega_K H} =  m_{s,turb} + m_{s, jet} = \alpha_v \varepsilon + 2 q \mu
,\end{equation}
\noindent where $c_s = \Omega_K H$ is the sound speed in the disk, and $m_{s,turb}$ and $m_{s, jet}$ are the contribution from the turbulent torque and jet torque, respectively. The expression for the turbulent torque $m_{s,turb} =  \alpha_v \varepsilon $ arises naturally in alpha disk theory \citep[see, e.g.,][]{SS73}. It is usually multiplied by a factor on the order of unity that depends only on the radial distributions of pressure and angular velocity, which we incorporated within the dimensionless $\alpha_v <1$ parameter for simplicity. Thus, in the SAD mode region, accretion takes place at a very subsonic pace ($m_{s,turb} <<1$) if the disk is geometrically thin. In a JED, $m_{s, jet} = 2 q \mu$ where $q \simeq -B_\phi^+/B_z $ is the magnetic shear of the magnetic configuration and $B_\phi^+$ is the toroidal magnetic field at the disk surface \citep[see][for more details]{Ferreira97}. We note that $q$ differs for each MHD solution so that $m_s$ and $\mu$ can be seen as independent within the allowed parameter range. The precise value of $m_{s, jet}$ depends on the trans-Alfv\'enic constraint, but accretion in a JED is always at least sonic and usually supersonic with $m_{s,jet} \gtrsim 1$ \citep{Ferreira97}. This is the reason why the two accretion modes are mutually exclusive and have different physical behaviors.

These JED dynamical parameters, that is, ($\mu, \xi, m_s, b$), are all related to each other for a given cold MHD solution. However, as shown for instance in \citet{Casse00b}, incorporating thermal effects at the disk surface introduces an additional degree of freedom. This is also the case with the source of the magnetic field diffusion \citep{Bethune17}. As a consequence and for the sake of completeness of our study here, we use $\mu, \xi, m_s$ , and $b$ as reasonably independent and uniform in radius. To summarize, a JED configuration is described by 
\begin{itemize}
        \item disk magnetization, $0.1 \leq \mu \leq 0.8$ 
        \item disk ejection efficiency, $0.01 \leq \xi \leq 0.2$
        \item accretion Mach number, $ 0.5 \leq  m_s \leq 3$
        \item jet power fraction, $0.2 \leq b \leq 0.8,$
\end{itemize}
where the range used for the dynamical parameters ($\mu, ~ \xi, ~ m_s, ~ b$) does not rely on any proxy or assumptions since it has been taken to be fully consistent with former MHD self-similar accretion-ejection calculations \citep[see][and references therein for more details]{Ferreira97, Casse00b, Petrucci10}.

\subsection{Computing the thermal structure of the local disk}
\label{sec:DiskConfig}

We assume a two-temperature $T_e \neq T_i$, fully ionized plasma with density $n_e = n_i$ and $\rho_0= m_i n_e$, $m_i$ being the proton mass. Indices $i$ and $e$ refer to ions and eletrons
throughout the paper. The electron density is related to the local disk accretion rate by
\begin{eqnarray}
n_e &=& \frac{\rho_0}{m_i} = \frac{\dot M}{4\pi m_i \Omega_K R^3} \frac{1}{m_s \varepsilon^2} =  n_* \frac{\dot{m}(r) r^{-3/2}}{\varepsilon^2 m_s } \nonumber \\
n_* &=& \frac{1}{\sigma_T R_g} \simeq 1.02 \times 10^{19} \, m^{-1}~~ \text{cm$^{-3}$}
\label{eq:density}
,\end{eqnarray}
where $\sigma_T$ is the Thomson cross section. Combining the disk quasi-static vertical equilibrium with the equation of state leads to
\begin{equation}
P_{tot}= n_e (k_BT_e + k_BT_i) + P_{rad} = \rho_0 \frac{GM}{R} \varepsilon^2 = P_*  \frac{\dot{m}(r) r^{-5/2}}{m_s } \label{eq:vert}
,\end{equation}
where $k_B$ is the Boltzmann constant and $P_*= m_i n_* c^2$. This equation provides a link between four independent variables: the disk aspect ratio $\varepsilon=h/r,$ and $T_e, T_i, \text{and
} P_{rad}$. 

The electron and proton temperatures are solved at each radius using the coupled steady-state local energy balance equations
  \begin{eqnarray}
     \left( 1-\delta \right) \cdot q_{turb} &=&  q_{adv,i} + q_{ie} ~~~~~~~~~~~~~~~~~~~~ \text{ions} \label{eq:IONS} \\
      \delta \cdot q_{turb} &=& q_{adv,e} - q_{ie} + q_{rad} ~~~~~~~~~ \text{electrons} \label{eq:ELECTRONS}
   ,\end{eqnarray}

\noindent where  $q_{turb}$ is the turbulent heating term, $q_{adv,i}$ ($q_{adv,e}$) the ion (electron) advection of internal energy, $q_{ie}$ the Coulomb collisional interaction between ions and electrons, and $q_{rad}$ the radiative cooling due to electrons. In principle, the released turbulent energy could be unevenly shared between ions and electrons by a factor $\delta$. We choose $\delta=0.5$, however, as discussed in the introduction, since JEDs require strongly magnetized plasmas. The expressions for the various terms are detailed in sections \ref{sec:Turb} to \ref{sec:BELM}.

\subsubsection{Turbulent heating}
\label{sec:Turb}

The exact expression of the turbulent heating term remains unknown since no precise theory of MHD turbulence is available. Instead, one relies on the disk angular momentum conservation equation since in a near-Keplerian disk, all available energy is stored as rotation. We use the following expression from \citet{Petrucci10}
\begin{eqnarray}
q_{turb} & = &  (1-b) (1-\xi) \frac{G M \dot{M}(R)}{8 \pi H R^3}  \nonumber \\
& \simeq & 1.5  \times 10^{21} (1-b) (1-\xi) \frac{\dot{m}(r)}{\varepsilon m^2 r^4}  ~~ \text{erg/s/cm$^{3}$}
\label{eq:HEAT}
.\end{eqnarray}
The power density heating term is a local decreasing function of the variable $\varepsilon(r)$ only, it is independent of other physical parameters such as $T_e$ or $T_i$.

\subsubsection{Coulomb interaction}

The electron-ion collisional coupling term is taken from \citet{Stepney83} 
\begin{eqnarray}
  q_{ie} &=& \frac{3}{2} \frac{m_e}{m_i}n_e n_i \sigma_T c \ln \Lambda (k_BT_i - k_BT_e) \Delta_{ie}  \nonumber \\
&\simeq  &  7.82  \times 10^{7} ~ \frac{\dot{m}(r)^2   r^{-3} }{m^2 \epsilon^4 m_s^2} \, \left ( T_i - T_e \right )   \Delta_{ie}  ~~ \text{erg/s/cm$^{3}$}  \label{eq:COULOMB} \\
 \Delta_{ie} &=& \frac{1}{K_2 ( 1/\theta_e ) K_2 ( 1/\theta_i )} \nonumber \\
  &~& \times \left[ \frac{2 (\theta_e+\theta_i)^2+1}{\theta_e+\theta_i} K_1 \left( \frac{\theta_e + \theta_i}{\theta_e \theta_i} \right) + 2 K_0\left(\frac{\theta_e + \theta_i}{\theta_e \theta_i}\right)\right]  \nonumber
\end{eqnarray} 
 \noindent where the temperatures are expressed in Kelvin, $m_e$ the electron rest mass, $\ln \Lambda = 15$ the Coulomb logarithm, $\theta_e= k_B T_e/m_e c^2$, $\theta_i= k_B T_i/m_i c^2$ and $K_{0/1/2}$ are the modified Bessel functions. The collisional term is thus a function of $T_i, T_e$ , and $\varepsilon$.

\subsubsection{Advection}
\label{sec:ADV}

For a fluid of a species $\alpha$ (electrons $e$, ions $i,$ or photons $rad$) with pressure $P_\alpha$, internal energy density $U_\alpha= P_\alpha/(\gamma_\alpha -1)$ with $\gamma_\alpha$ the adiabatic index and velocity $\vec u$, the advection term writes  
\begin{eqnarray}
 q_{adv,\alpha} &= & P_\alpha \text{div} \vec{u} + \text{div} U_\alpha \vec{u} = \frac{\gamma_\alpha P_\alpha}{\gamma_\alpha-1} \text{div} \vec{u} + \frac{\vec u \cdot \nabla P_\alpha}{\gamma_\alpha-1} \nonumber \\
&=& \frac{u_R P_\alpha}{R} \left (\frac{\gamma_\alpha}{\gamma_\alpha-1} \left (1 + \frac{\text{d} \ln u_R}{\text{d} \ln R} \right ) + \frac{1}{\gamma_\alpha-1} 
 \frac{\text{d} \ln P_\alpha}{\text{d} \ln R}  \right ) 
,\end{eqnarray}
where the last expression uses the fact that the only relevant speed is the radial (accretion) component. This can be further simplified to 
\begin{eqnarray}
 q_{adv,\alpha} &= & - m_s \varepsilon \Omega_K P_\alpha \Delta_\alpha \nonumber \\
\Delta_\alpha &=& \frac{\gamma_\alpha}{\gamma_\alpha-1} \left (\frac{1}{2} + \frac{\text{d} \ln m_s \varepsilon}{\text{d} \ln r} \right ) + \frac{1}{\gamma_\alpha-1}  \frac{\text{d} \ln P_\alpha}{\text{d} \ln r} \label{eq:ADV_deltaalpha}
.\end{eqnarray}

For the sake of simplicity, we assume that ions and electrons have the same accretion speed (same $m_s$) and that electrons are relativistic ($\gamma_e=4/3$), whereas ions remain non-relativistic ($\gamma_i=5/3$). Moreover, the radiation pressure $P_{rad}$ and radiation energy density $U_{rad}$ are assumed to follow the law $U_{rad}= 3 P_{rad}$ ($\gamma_{rad}=4/3$), which is strictly valid only in optically thick media. However, if the plasma is optically thin, radiation leaves the system and is expected to have very little effect on energy advection. The error introduced is therefore negligible. We finally obtain
\begin{eqnarray}
 q_{adv} &= &  q_{adv,i} + q_{adv,e} +q_{adv,rad} \nonumber \\
 &= & q_{adv,*} \left (\frac{P_i}{P_{tot}} \Delta_i  + \frac{P_e}{P_{tot}} \Delta_e + \frac{P_{rad}}{P_{tot}} \Delta_{rad} \right )   \label{eq:ADV}  \\
 q_{adv,*}  &= & - m_s \varepsilon   \Omega_K P_{tot} = - \varepsilon \frac{G M \dot M(R)}{4 \pi R^4} \nonumber \\  
 &=& - 3\times 10^{21} \frac{\epsilon \dot{m}}{m^2 r^{4}}~~ \text{erg/s/cm$^{3}$}  \nonumber
\end{eqnarray}
where $P_i= n_i k_BT_i$ and $P_e= n_e k_B T_e$. These two terms do not only depend on the variables $\varepsilon, T_e, T_i$ ,
and $P_{rad}$, but also on their radial derivatives. This introduces numerical complications that require caution (see Appendix A for more details). In a usual dense cold disk, the $\Delta_\alpha$ factors are on the order of unity and $q_{adv}/q_{turb} \propto \varepsilon^2 / (1-b)$, implying a negligible contribution from advection to the disk energy balance when it is cold ($\varepsilon<<1$).

\subsubsection{Radiation}
\label{sec:BELM}
In our study, radiation is treated in a one-zone approach. The thermal balance of the disk is solved at each radius $R$ and the corresponding self-consistent continuum emission is obtained by summing the contribution of each disk annulus of radius $R$, width $dR$ and half-thickness $H(R)$. The continuum emission of accretion flows results from Compton scattering and emission/absorption through bremsstrahlung and synchrotron radiation. The radiation pressure $P_{rad}$ and the bolometric radiation cooling term $q_{rad}$ are described by single bridge functions that can accurately handle the optically thin and optically thick regimes, as described below.  
\bigskip

\noindent \underline{Optically thick radiation} \\

Optically thick solutions are well described within the radiation diffusion limit. The local emission $dL_{\nu}^{(thick)}$ from a disk annulus is then a simple blackbody with effective temperature $T_{eff} = (4/3\tau_{tot})^{1/4}T_e$, where $T_e$ is the disk mid-plane temperature and $\tau_{tot} $ the total (half-) disk optical depth (defined below). The associated cooling rate is
\begin{equation}
q_{thick} = \frac{1}{H} \sigma T_{eff}^4 = \frac{1}{H} \frac{4 \sigma T_e^4}{3 \tau_{tot}} \label{eq:qbb}
,\end{equation}
where $\sigma$ is the Stefan-Boltzmann constant.
\bigskip

\noindent \underline{Optically thin radiation} \\

The optically thin emission is computed using \textsc{BELM} \citep{Belmont08, Belmont09}. This code computes the emission from a steady, uniform, spherical, magnetized cloud using the exact cross sections and including both synchrotron and bremsstrahlung radiation and thermal Compton scattering. Although it was designed to handle complicated, self-consistent particle distributions, we enforce here the use of thermal particle distributions. Each disk annulus is decomposed into $dN$ spheres of radius $H(R)$ filled with a plasma of density $n_e$ (or Thomson optical depth $\tau_T = \sigma_T n_e H$), electron temperature $T_e$ , and magnetic field $B$. For simplicity, we assume a purely vertical field $B=B_z$ such that
\begin{equation}
B_z = B_* \left ( \frac{\mu}{m_s}\right )^{1/2}\dot m^{1/2} r^{-5/4}
,\end{equation}
where $B_*= \sqrt{\mu_o P_*}$, consistent with the JED dynamical mode. To summarize, injecting these plasma parameters into the \textsc{BELM} code provides at each radius (1) the optically thin cooling rate  $q_{thin}$ and (2) the corresponding spectrum $l_\nu$ emitted by a sphere. The total optically thin spectrum emitted by the disk annulus is simply $dL_{\nu}^{(thin)}= dN l_\nu$, where $dN=4\pi R H dR/(4\pi H^3/3)$. For more details, we refer to Appendix A and the papers on the \textsc{BELM} code. 
\bigskip

\noindent \underline{Transition from thin to thick} \\

The thin and thick cooling expressions are then inserted into a general bridge formula that is valid for all regimes \citep{Hubeny90, Chen95, Artemova96, Esin96}. Defining the absorption, effective, and total optical depths as  
\begin{eqnarray}
\tau_a &=& \frac{2}{3\tau_{tot} } \frac{q_{thin}}{q_{thick}} \\
\tau_*  &=& (3 \tau_{tot} \tau_a/2)^{1/2}\\
\tau_{tot} &=& \kappa_R \rho_0 H 
,\end{eqnarray}
respectively, where the total Rosseland mean opacity $\kappa_R=\kappa_T+\kappa_{ff}$ is computed using the Thomson $\kappa_T= \sigma_T/m_i$ and the free-free $\kappa_{ff}= 5 \times10^{24} \rho_o T_e^{-7/2} \, $cm$^2/$g opacity laws, the total cooling rate can be approximated as
\begin{equation}
q_{rad} = q_{thick} \left( 1 + \frac{4}{3 \tau_{tot}}  + \frac{e^{- \frac{\tau_{tot}}{100}}}{\tau_*^2} \right)^{-1} \label{eq:rad}
.\end{equation}
The numerical coefficients vary from one reference to another but are not expected to produce significant deviations. Compared to previous work, an exponential coefficient $e^{- \frac{\tau_{tot}}{100}}$ was also added to enforce the optically thick-thin transition at very high Thomson optical depth, as in \citet{Wandel91}. Similarly, the radiation pressure is estimated in the gray and Eddington approximation by
\begin{equation}
P_{rad} = \frac{q_{rad} H }{c} \left( \tau_{tot} + \frac{4}{3} \right)  \label{eq:Prad}
\end{equation}
and can be computed at any disk optical depth as a function of $n_e$, $T_e$, $\mu$, and $\varepsilon$. 

Consistently with the energy balance equation (\ref{eq:rad}), a bridge formula is also required to compute the spectrum $dL_\nu$ emitted by each annulus. For this purpose, we use
\begin{eqnarray}
 dL_{\nu} = (1-a) dL_{\nu}^{(thick)}  + a dL_{\nu}^{(thin)}  \label{eq:Fnu}
,\end{eqnarray}
where $a = e^{-\tau_*^2}/ (e^{-\tau_*^2} + e^{-1/\tau_*^2})$, such that $dL_{\nu}= dL_{\nu}^{(thick)}$ when $\tau_* \gg1$ and $dL_{\nu}= dL_{\nu}^{(thin)}$ when  $\tau_* \ll1$. The total disk spectrum is computed by integrating over the entire disk $L_\nu = \int dL_\nu$ and the total received flux $ F_\nu$ is computed assuming a face-on object located at a distance $D$, namely $F_\nu= L_\nu/4\pi D^2$. In order to be specific and display spectra in physical units, we focus in this paper on GX 339-4, with $D = 8 \pm 4$~kpc \citep[see][]{Miller04} and a mass $m = 5.8 \pm 0.5$ \citep[see][]{Hynes03}, although this source is not entirely face-on, as suggested in different work \citep{Cowley02, Miller04, Gallo04}.

\section{JED thermal states} \label{sec:JEDthermalstates}

For a set of disk parameters and at any given radius $r$, the set of equations (\ref{eq:vert}), (\ref{eq:IONS}), (\ref{eq:ELECTRONS}), (\ref{eq:rad}), and (\ref{eq:Prad}) allows solving for the variables $\varepsilon, T_e, T_i,$ and $P_{rad}$, providing thereby the radial distributions for these quantities as well as the emitted spectrum.
The main difficulty is the nonlocal nature of the advection term since it depends on the radial derivatives of both the ion and electron temperatures. Commonly, either an assumption about the value (and sign) of its logarithmic derivative \citep[e.g.,][]{Narayan96, Oda10} or the computation of a global transonic flow \citep[e.g.,][and references therein]{Yuan00, Artemova06} have been used. Here, we start  at $r_{out}$ and progress inward down to $r_{in}$, keeping track of the previous thermal solution and obtaining thereby a consistent advection term. The difficulty is amplified because very often, three solutions can be found at a given radius \citep[as shown in][]{Petrucci10}. In this entire section, we use only $15$ to $20$ radial steps, but for section \ref{sec:Hysteresis} and \ref{sec:bestparams}, the highest resolution has been used so that the spectra and the thermal disk structure are no longer affected by the radial discretization. For more details on this and on our numerical methods, see Appendix \ref{sec:Method}.

%
\subsection{Three different JED solutions}
\label{sec:ExampleOfSolution}

As shown in \citet{Petrucci10} in their one-temperature JED calculations, there may be either only one or three thermal equilibrium solutions at any given radius. This is of course also verified in our two-temperature resolution.    
\begin{figure}[h!]
   \centering
   \includegraphics[width=\columnwidth]{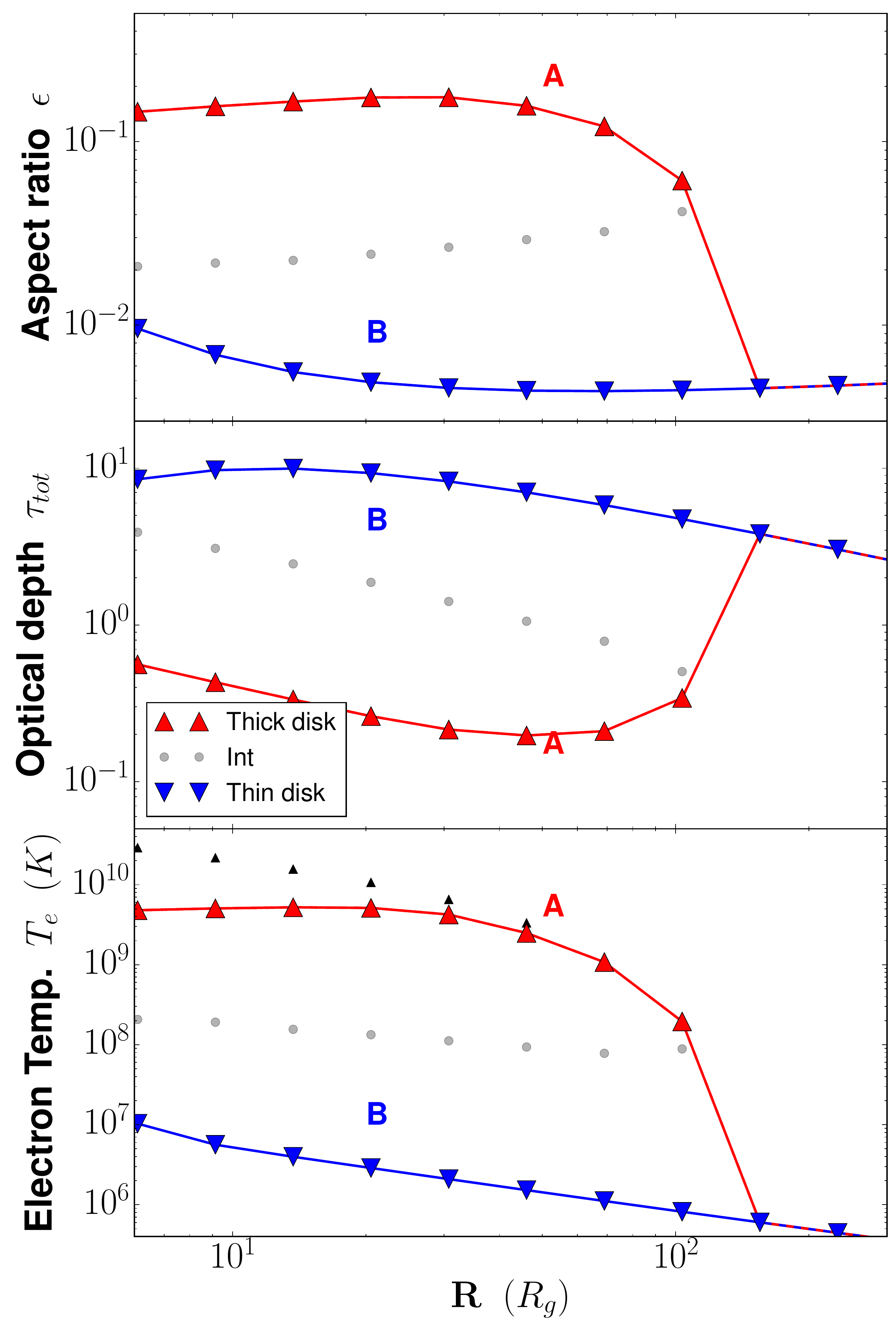}
      \caption{JED configuration with $\dot{m}_{in} = 0.3$ and ($\mu= 0.1,~ \xi= 0.02,~ m_s=1.5,~ b=0.7$). The top panel shows the disk aspect ratio $\varepsilon =h/r$, the middle panel presents the total optical depth $\tau_{tot}$ , and the bottom panel shows the electron temperature $T_e$ in K. The temperature $T_i$ of the ions, when different from that of the electrons, is also shown as black markers. In these three panels, top-oriented triangles indicate the optically thin, geometrically thick (hot) solution, and  bottom-oriented triangles indicate optically thick, geometrically thin (cold) solution. When present, the unstable intermediate solution is shown as gray dots. The two possible JED configurations A and B are defined as the red and blue distributions (see text).}
         \label{fig:Exple}
\end{figure}

We first examine a JED settled from $r_{in}=6$ to $r_{out}= 5\times 10^2$ with $\dot{m}_{in} = 0.3$ and dynamical parameters ($\mu= 0.1,~ \xi= 0.02,~ m_s=1.5,~ b=0.7$). The solution of the thermal equilibrium for all disk radii is shown in Figure~\ref{fig:Exple}. This figure displays the radial distributions for the optical depth of the entire disk $\tau_{tot}$, electron central temperature $T_e$ (and $T_i$ when different), and disk aspect ratio $\varepsilon$. In the outer parts of the disk, down to a transition radius $r_t \simeq 180$, there is only one possible solution. The disk is in thermal balance with (mostly) radiation diffusion balancing turbulent heating, leading to a cool, optically thick and geometrically thin disk. Below this transition radius and down to the innermost disk radius $r_{in}$, however, three steady-state solutions are
possible: (1) the same cold/geometrically thin as for $r > r_t$, (2) a much hotter, geometrically thick and  optically thin (with $T_e < T_i$) solution, and (3) an intermediate solution that is thermally unstable \citep[see][and references therein]{Petrucci10} and is therefore not considered further in this work. Each of the two stable solutions is described in greater detail below. 

The two different thermally stable solutions, the thick/hot disk and the thin/cold disk, are quite generic \citep{Yuan14, Petrucci10} and appear in a range of disk radii for many parameter sets. Therefore, when making a model describing an outbursting object, for instance, one needs to decide which of these local solutions will be used. In our code we assume that any physical modification in accretion rate is stimulated by the outer parts of the disk. Thus, advection is self-consistently taken into account using outside-in calculations. This assumption implies that once the system "chooses" one of the two solutions at the transition radius $r_t$, it stays there as long as its existence remains possible. In a way, the advection term carries the memory of the outer solution to the inner radii, thereby enforcing a continuity. This memory effect allows two possible JED configurations for the same parameter set, defined as A and B. They are shown as red and blue lines in Fig.~\ref{fig:Exple}. 

\begin{figure}[h!] 
   \centering
   \includegraphics[width=\columnwidth]{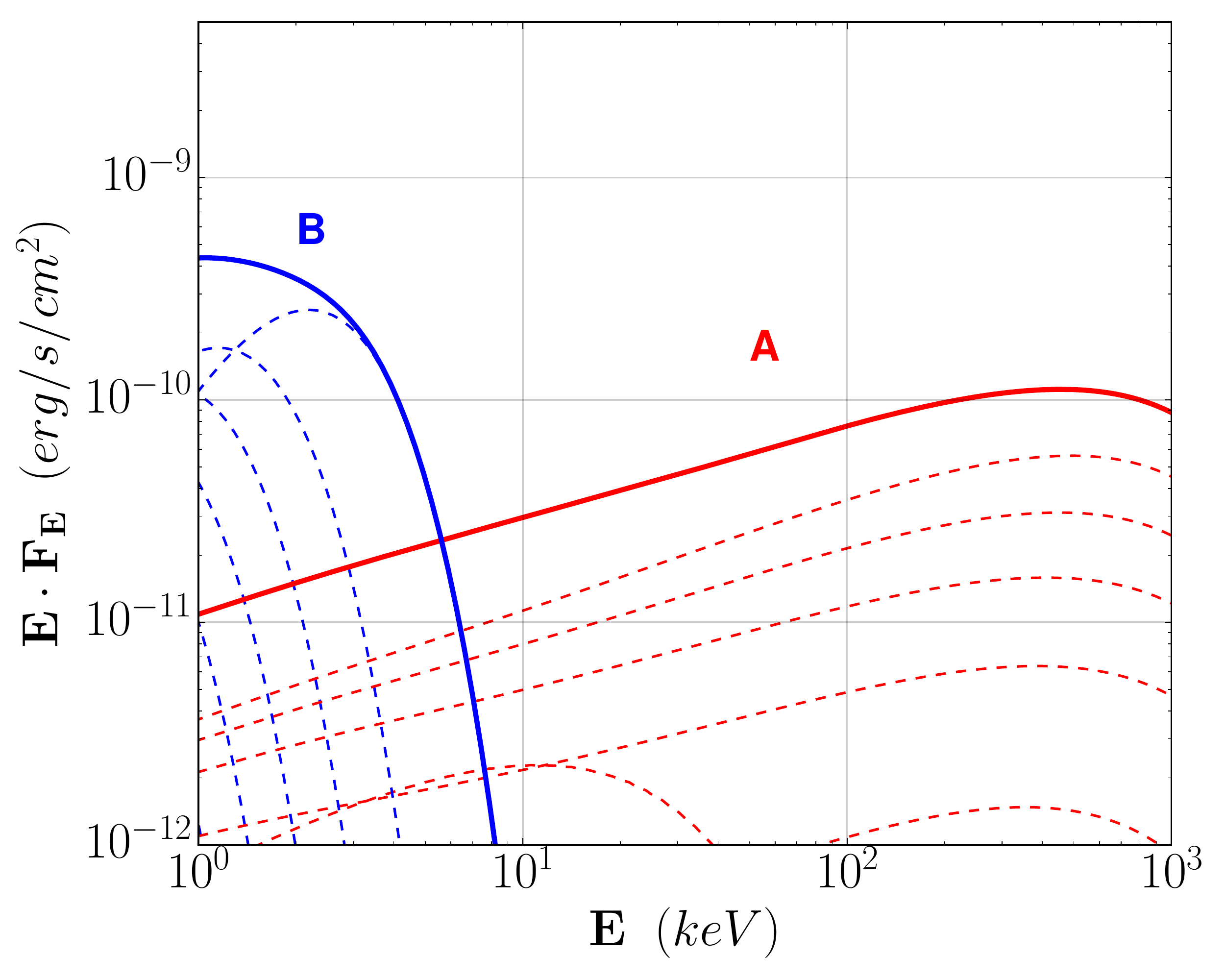}
      \caption{Spectral energy distributions for the JED configurations A (red) and B (blue), as defined in Fig. \ref{fig:Exple}. The dashed lines correspond to local spectra produced at individual radii for which a solution has been computed (highlighted by triangles in Fig. \ref{fig:Exple}). The thick solid line is the total spectrum from the whole accretion disk (\textup{i.e., }the integration over all radii).}   
         \label{fig:Exple_sed}
\end{figure}

In JED configuration A, the disk switches to a much hotter and geometrically thicker disk. Because the transition radius is quite large ($r_t= 180$), this configuration is spectrally dominated by the inner hot solution. It is characterized by a typical electron temperature $T_e \sim3\times 10^{9}$ K, usually lower than the ion temperature $T_i$  (see black triangles in the bottom panel
of Fig. \ref{fig:Exple}) because its density is very low ($\sim 10^{5-6}$ times lower than in a typical SAD mode) and Coulomb interactions cannot entirely thermalize the disk ($T_e \neq T_i$). It is supported by gas pressure ($P_{gas} \gg P_{rad}$) and is optically thin ($\tau_{tot} \lesssim 1$) and geometrically thick ($\varepsilon \gtrsim 0.1$). The dominant cooling term is advection with $q_{adv} > q_{rad}$ , and the spectrum is a typical Comptonized emission that can be fitted by a simple power law with an exponential cutoff: $\nu F_\nu = E F_E\propto E^{2-\Gamma} e^{-(E/E_{cut})}$.  For the current set of parameters, we obtain a typical photon spectral index $\Gamma \simeq 1.5-1.6$  and a high-energy cutoff $E_{cut} > 100$ keV (red curves in Fig.~\ref{fig:Exple_sed}). 

Conversely, JED configuration B describes  a disk remaining marginally optically thick and geometrically thin all the way down to the innermost orbit. This solution is characterized by a gas-supported disk with a typical central temperature of up to $2\times 10^7$ K, marginally optically thick ($\tau_{tot} >10$), and geometrically thin ($\varepsilon \sim 0.01$). This solution is similar to the \citet{SS73} solution, but $\sim 10^{2-3}$ times less dense and with a lower central temperature since a fraction of the released energy is carried away by the jets. 
Figure~\ref{fig:Exple_sed} displays the spectrum emitted by this configuration (blue curves). Each disk annulus emits mostly as a blackbody (dashed lines) of effective temperature $T_{eff}(r)$ so that the global disk spectrum (solid line) is indeed a multicolor-disk
blackbody with an internal temperature $T_{eff,in} = 1.72$~keV for the chosen parameters.

\subsection{New solution: the slim JED}
\label{sec:Slim}

\begin{figure}[t]
   \centering
   \includegraphics[width=\columnwidth]{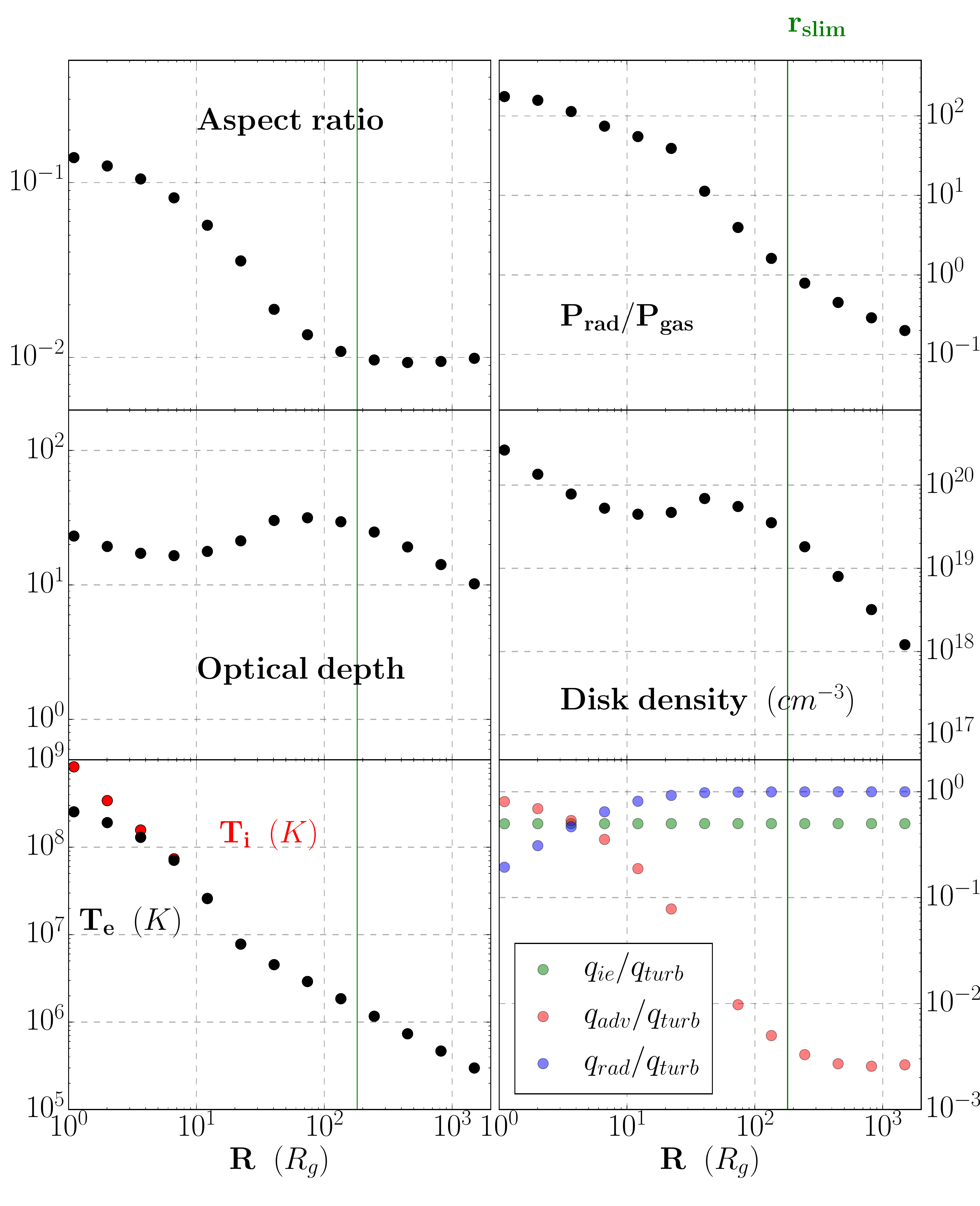}
      \caption{Slim JED configuration settled from $r_{in} = 1$ to $r_{out}= 2\times 10^3$ with  $\dot{m}_{in} = 5$,  obtained for ($\mu= 0.1,~ \xi= 0.02,~ m_s=1.5,~ b=0.7$). From left to right, top to bottom, are shown the following radial distributions: disk aspect ratio $\varepsilon = h/r$, radiation to gas pressure ratio $P_{rad}/P_{gas}$, total optical depth  $\tau_{tot}$, electron density $n_e$ (in cm$^{-3}$), electron $T_e$ (black) and ion $T_i$ (red) temperatures in K, relevant cooling ($q_{rad}, q_{adv}= q_{adv,e}+q_{adv,i}$) processes with respect to heating $q_{turb}$ in erg/s/cm$^3$.}
         \label{fig:Slim}
\end{figure}

As for the SAD mode, the JED mode can also exhibit thermal properties that resemble the SLIM disk structure \citep[see seminal work from][]{Abramowicz88}, namely an optically and geometrically thick disk where advection of internal energy provides a significant if not dominant cooling term. The existence of a slim solution is found to be possible only at high accretion rates, but it is favored in a JED with respect to the SAD mode because of the lowest density at the same accretion rate. Slim JED solutions are found for $\dot{m}_{in} \gtrsim 0.5$, whereas one need to go as high as $\dot{m}_{in} = 10$ in the SAD mode, see figure \ref{fig:SADSlim}, in the innermost disk regions ($r < 10-100$) where the disk is optically thick and dominated by the radiation pressure ($P_{rad} \gtrsim P_{gas}$).

\begin{figure}[h!]
   \centering
   \includegraphics[width=\columnwidth]{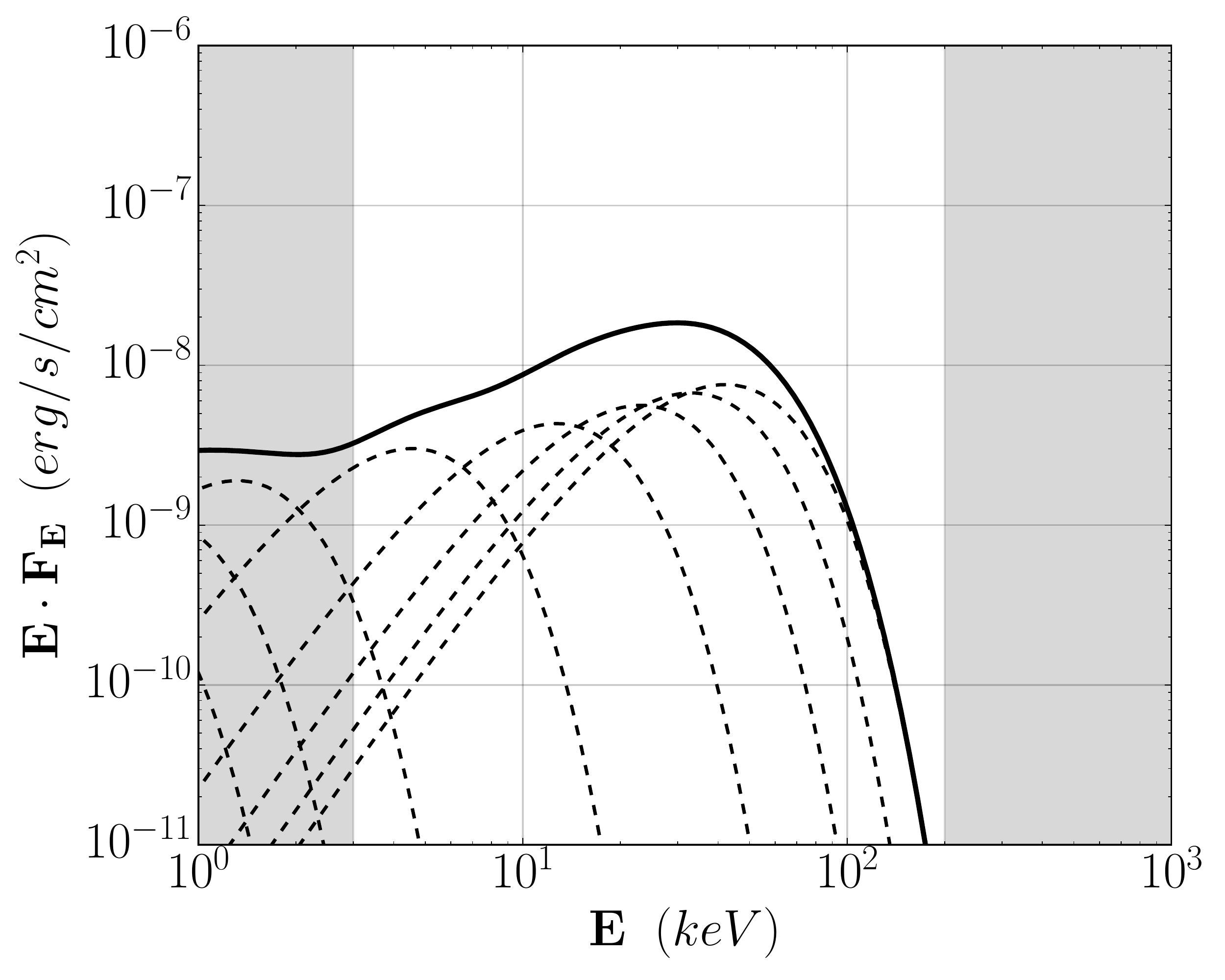}
      \caption{Spectral energy distribution of the slim configuration shown in Fig.~\ref{fig:Slim}. Each dashed line is the spectrum emitted by the annulus located at one of the dots appearing in Fig.~\ref{fig:Slim}, the solid line being the total spectrum, and the grey areas are beyond the scope of the spectral studies we dispose: \textsc{Hexte} \& \textsc{Pca} spectral band, \textit{i.e.} $E \in [3, ~ 200]$ keV.}
         \label{fig:Slim_sed}
\end{figure}

Figure \ref{fig:Slim} provides an example of such a slim configuration, settled from $r_{in} = 1$ to $r_{out}= 2\times 10^3$ with  $\dot{m}_{in} = 5$ and obtained with the dynamical parameters  ($\mu= 0.1,~ \xi= 0.02,~ m_s=1.5,~ b=0.7$). The radial distributions of various disk quantities are shown from the upper left to the bottom right
panel: the disk aspect ratio $\varepsilon=h/r$, total optical depth $\tau_{tot}$, electron $T_e$ and ion $T_i$ temperatures, ratio of radiation to gas pressure $P_{rad}/P_{gas}$, disk density $n_e$ , and several cooling/heating terms. At radii larger than $r \simeq 200$, the disk is supported by the gas pressure, and its thermal state resembles that of a Shakura \& Sunyaev disk, namely a geometrically thin $\varepsilon \sim 10^{-2}$, thermalized and rather cold ($T_e = T_i \sim 10^5-10^6$ K) dense plasma with $n_e \sim 10^{18}-10^{19} ~ cm^{-3}$. Advection is negligible with  $q_{adv} / q_{rad} \simeq 10^{-3}-10^{-2}$ , and given an optical depth higher than unity ($\tau_{tot} > 10$), the disk cools down mostly through its blackbody radiation. 
Below  $r \simeq 200$, however, radiation pressure takes over the gas pressure and the inner disk characteristics are deeply modified. The disk aspect ratio rises by a factor 10 to reach $\varepsilon \sim 0.2$, and the optical depth $\tau_{tot}$ and disk density $n_e$ stop their rising path to reach more constant values ($\tau_{tot} \simeq 10-20$ and $n_e \simeq 10^{20} ~ cm^{-3}$). The disk is no longer thermalized, electron and ion temperatures reach $T_i \sim 3 T_e \sim 2 \times 10^8$ K, and advection becomes a major cooling process with $q_{rad} \simeq 0.2 q_{adv}$. The corresponding spectrum is shown in Fig.~\ref{fig:Slim_sed}. Since the disk is optically thick at all radii, it is the sum of blackbodies (dashed lines) but with an effective temperature radial exponent $q \neq3/4$ due to the dominant influence of advection. In the \textsc{Hexte} \& \textsc{Pca} spectral band ($E \in [3, ~ 200]$ keV), the resulting spectrum mimics a \textit{\textup{hard-state}} spectrum with a photon index $\Gamma \simeq 1.2$ and a high-energy cutoff $E_{cut} < 100$ keV. We show in section \ref{sec:bestparams} that these slim states may have a drastic observational importance.

\section{Influence of the JED dynamical parameters}
\label{sec:Params}

Our goal is to achieve a simplified physically motivated picture of the innermost regions of XrBs in all their stages. In this subsection, we investigate the effect of the main parameters, namely, $b$, $\mu$, $r_{in}$, $m_s$ , and $\xi$. Although crucial, the role of the accretion rate is studied in the following section and especially in a forthcoming companion paper.

\begin{table}[h!]
\centering
\begin{tabular}{c c c c c c c c}
\hline \hline
Fig. & $\dot{m}_{in}$ & $r_{in}$ & $\mu$ & $\xi$ & $m_s$ & $b$ \\ \hline
\ref{fig:EffectOfB} & $0.5$ & $3$ & $0.1$ & $0.1$ & $1.5$ & $0.3, ~ 0.7$ \\
\ref{fig:EffectOfMu} & $0.5$ & $3$ & $0.1, ~ 1$ & $0.1$ & $1.5$ & $0.5$ \\
\ref{fig:EffectOfRi} & $0.08$ & $1, ~ 5$ & $0.1$ & $0.3$ & $1.5$ & $0.5$ \\
\ref{fig:EffectOfRi_OnSlim} & $0.08$ & $1 \rightarrow 6 $ & $0.1$ & $0.1$ & $1.5$ & $0.3$ \\
\ref{fig:EffectOfMs} & $0.5$ & $3$ & $0.1$ & $0.1$ & $0.75, ~ 1.5$ & $0.5$ \\
\ref{fig:EffectOfXi} & $0.6$ & $3$ & $0.1$ & $0.01, ~ 0.2$ & $1.5$ & $0.5$ \\ \hline
\end{tabular}
\caption{Parameter sets used in this section for each figure.}
\label{table:ParamSets}  
\end{table}

The parameter space is quite large, and it can safely be expected that different combinations lead to comparable observational signatures. Removing this degeneracy might be possible when considering the time-dependent disk dynamics, but this is beyond the scope of our present study. Instead, we only show here the effects of a given parameter when we move away from a fiducial JED configuration defined by $r_{in} = 3$ and the set ($\mu= 0.1,~ \xi= 0.1,~ m_s=1.5,~ b=0.5$), while we still have the same parameters as GX 339-4 ($D = 8$ kpc and $m = 5.8$) and a variable accretion rate $\dot{m}_{in}$. The parameter sets chosen to illustrate the effect of the JED dynamics are detailed in Table~1. 

\subsection{Jet power fraction $b$}

\begin{figure}[ht]
  \centering
  \includegraphics[width=\columnwidth]{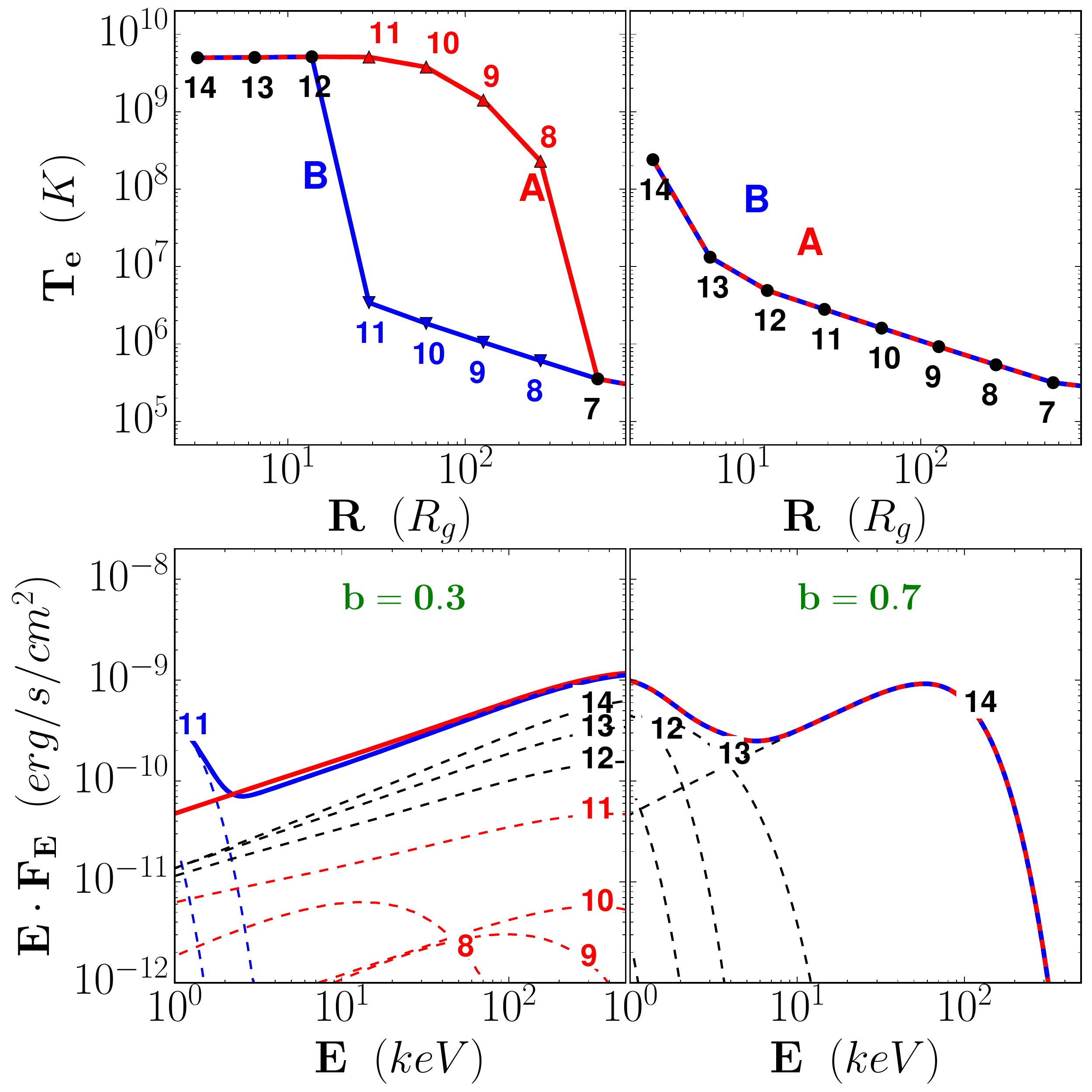}  
  \caption{Effect of the jet power fraction $b$ on a JED configuration fed with $\dot{m}_{in} = 0.5$ at $r_{in}=3$, $b = 0.3$ (left) and $b = 0.7$ (right). The top panels show the electron temperature, and the bottom panels display the local spectra emitted by each radius (numbered dashed lines) and the corresponding total disk spectrum (solid line), according to each configuration A (red) and B (blue). When configurations A and B are identical, blue-red dashed lines are used.}
  \label{fig:EffectOfB}
\end{figure}

The parameter $b$ is the fraction of the released accretion power that is carried away by the two jets. This power leakage has a direct effect on the disk thermal equilibrium because the turbulent heating is lower than within a SAD configuration. The JED luminosity is therefore lower than that of a SAD fed with the same $\dot m_{in}$. Figure~\ref{fig:EffectOfB} shows two different JED configurations fed with $\dot{m}_{in} = 0.5$, one with $b = 0.3$ (left), and the other with $b = 0.7$ (right).   $b$ clearly has two main effects. 

First, it determines the number of existing local thermal solutions (along with $\dot m_{in}$ , of course). When $b=0.3$, two thermally stable solutions are possible below $r \simeq 530$ (dot 7 in Fig.~\ref{fig:EffectOfB}), while for $b=0.7$, only the colder, geometrically thin solution is possible. This is very important, as the former case gives rise to two distinct configurations A and B (hence a possible hysteresis, see next section), while in the latter case, there is only one configuration. As $b$ decreases, the local heating term increases accordingly, leading to a hotter and thicker disk, which makes it easier for the advection term to compete with radiation. When $b$ is large, all the power goes to the jets and the JED is a quite cold, weakly dissipative structure. 

The second effect of $b$ can be seen on the temperatures reached by the cold thermal solutions present in both cases. For $b=0.3$, the turbulent heating is higher than for $b=0.7$ and the disk achieves higher temperatures. The $b = 0.3$ solution is able to reach the slim disk state here, which explains the huge spectral and temperature differences between the two configurations B. The bottom left spectrum shows that configuration B becomes progressively slim around the point number 12 (below $r \sim 10$). The global disk spectrum is thus dominated by these innermost radii, which leads to a spectral signature that is totally different from the multi-blackbody displayed in the right case. We note, however, that for our choice of parameters, the $b=0.7$ solution also
becomes slim at $r=r_{in}$ (point 14), as shown by the modified spectrum emitted by this annulus.

\subsection{Disk magnetization $\mu$}

\begin{figure}[h!]
  \centering
  \includegraphics[width=\columnwidth]{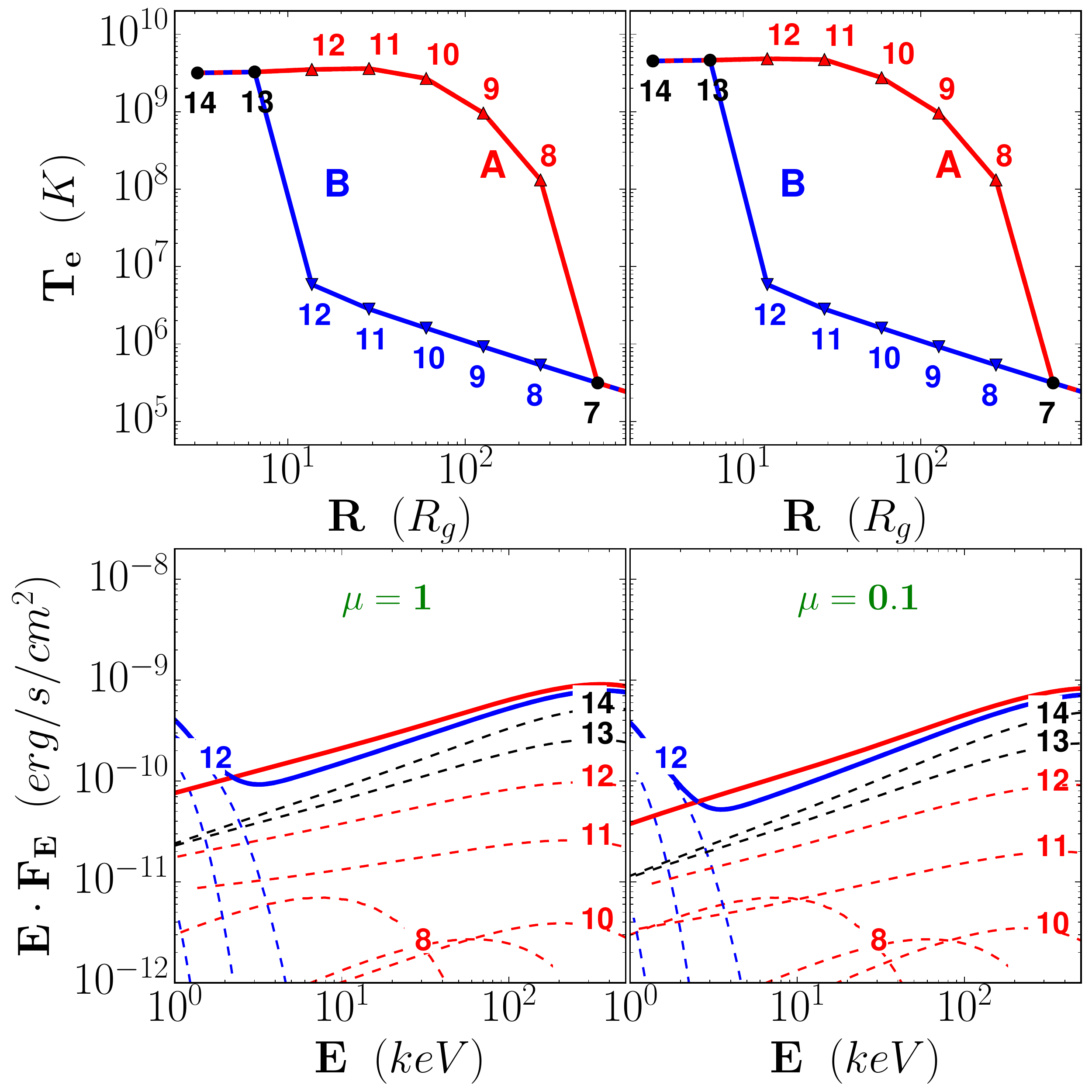}
  \caption{Effect of the disk magnetization $\mu$ on a JED configuration fed with $\dot{m}_{in} = 0.5$ at $r_{in}=3$, $\mu= 1$ (left) and $\mu = 0.1$ (right). This figure is similar to Fig.~\ref{fig:EffectOfB}.}
  \label{fig:EffectOfMu}
\end{figure}

We recall that the local mid-plane disk magnetization is $\mu = B_z^2/(\mu_o P_{tot})$, where $P_{tot}= P_{gas} + P_{rad}$.  We show in Fig.~\ref{fig:EffectOfMu} two different JED configurations fed with $\dot{m}_{in} = 0.5$, obtained for $\mu = 1$ (left) and $\mu = 0.1$ (right).

The disk magnetization has a much less drastic effect than $b$. Since it provides a scaling for the disk magnetic field, its main effect is on the radiation cooling efficiency and spectral shape through synchrotron emission when the disk is optically thin. As Fig.~\ref{fig:EffectOfMu} clearly shows, the optically thick solutions (configurations B) are not affected by the value of $\mu$. Even the optically thin solution (configuration A) is only barely affected, mostly because synchrotron emission peaks below $0.1$~keV. The influence is only visible for point 12 in the electron temperature, but the overall disk spectrum is almost identical for the two magnetization values\footnote{We note that this result is also dependent on the disk accretion rate used. For other values, varying $\mu$ could lead to higher or lower synchrotron emission so that an optically thin thermal balance would no longer be possible.}.

\subsection{Disk internal radius $r_{in}$}

\begin{figure}[h!]
  \centering
  \includegraphics[width=\columnwidth]{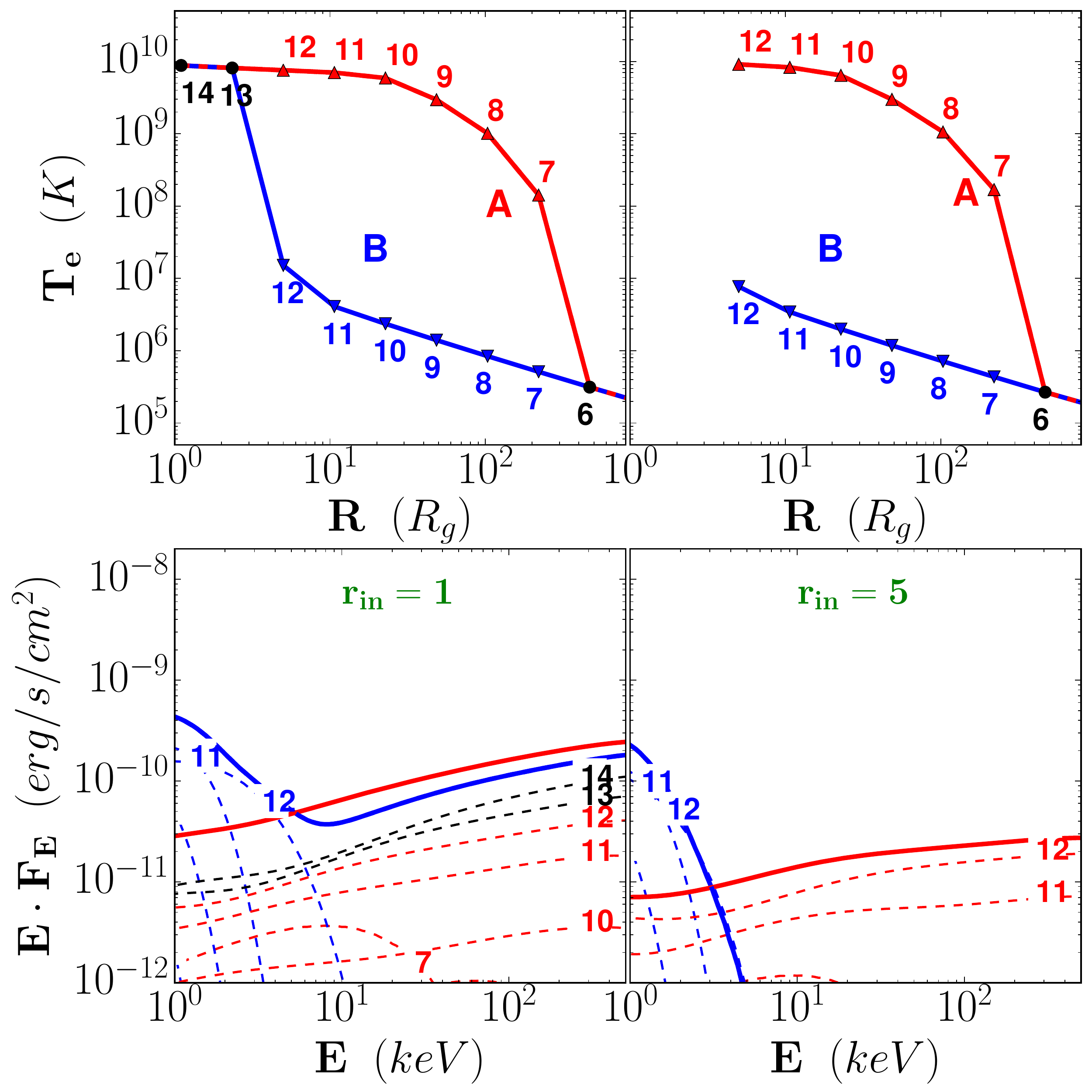}
  \caption{Effect of the disk innermost radius $r_{in}$ on a JED configuration fed with $\dot{m}_{in} = 0.08$ at $r_{in} = 1$ (left) or $r_{in} = 5$ (right). This figure is similar to Fig.~\ref{fig:EffectOfB}. Note that $\xi=0.3$ here to highten the effect of varying the inner radius.}
   \label{fig:EffectOfRi}
\end{figure}

In our simplified Newtonian approach, we use the disk innermost radius $r_{in}$ as a proxy for the black hole spin. This might vary from $r_{in} = 1$, in principle, for a maximally rotating Kerr black hole to $r_{in} = 6$ for a non-rotating Schwarzschild black hole. Thus, one would expect that the smaller the internal radius, the higher the disk luminosity, as the total available power scales as  $P_{acc} \propto r_{in}^{-1}$. Energy can also be advected onto the black hole, however, and does not necessarily need to be radiated, as seen in thick- or thick-disk states. To assess the effect of $r_{in}$ , we present in Fig.~\ref{fig:EffectOfRi} two different JED calculations for $r_{in} = 1$ (left) and $r_{in}=5$ (right), fed with  $\dot{m}_{in} = 0.08$, and $\xi = 0.3$ here to highten the effect of the inner radius on the accretion rate distribution.

As expected, for $r_{in}=1$ (left), the disk reaches deeper into the potential well and gives rise to a much higher luminosity, $L^{tot}_{r_{in}=1} = 2.1 \times 10^{-2} L_{Edd}$ and $L^{tot}_{r_{in}=5} = 7.9 \times 10^{-3} L_{Edd}$ for both configurations A. This is also somewhat enhanced because we defined the disk accretion rate at $r_{in}$. Changing the internal radius while keeping the same internal accretion rate within a JED configuration with $\dot m\propto r^\xi$ leads to a slight mismatch: $\dot m(r=5)= \dot m_{in}$ when $r_{in}= 5$ (right), while $\dot m(r=5)= \dot m_{in}(5/1)^\xi\simeq 1.17 \dot m_{in}$ when $r_{in}=1$ (left). Except for this slight discrepancy, the qualitative thermal behavior of the two solutions is different for configurations A and B. The hot A configuration is continued by a thick solution when  $r_{in}=1$, leading to higher radiation but with no qualitative difference in the spectrum with respect to $r_{in}= 5$. Conversely, the cold B configuration gives rise to a thick/thin solution below point 11 when $r_{in}= 1$, with a drastic influence on the overall disk spectrum. Only the existence of a blackbody below $2$~keV would allow distinguishing it from configuration A. This effect is of course amplified here by our choice of a high inner accretion rate. A smaller $\dot{m}_{in}$ would not lead to an inner slim thermal solution, and the effect on the overall disk spectrum would merely be a higher disk luminosity without changing its shape.
\begin{figure}[h!]
  \centering
  \includegraphics[width=\columnwidth]{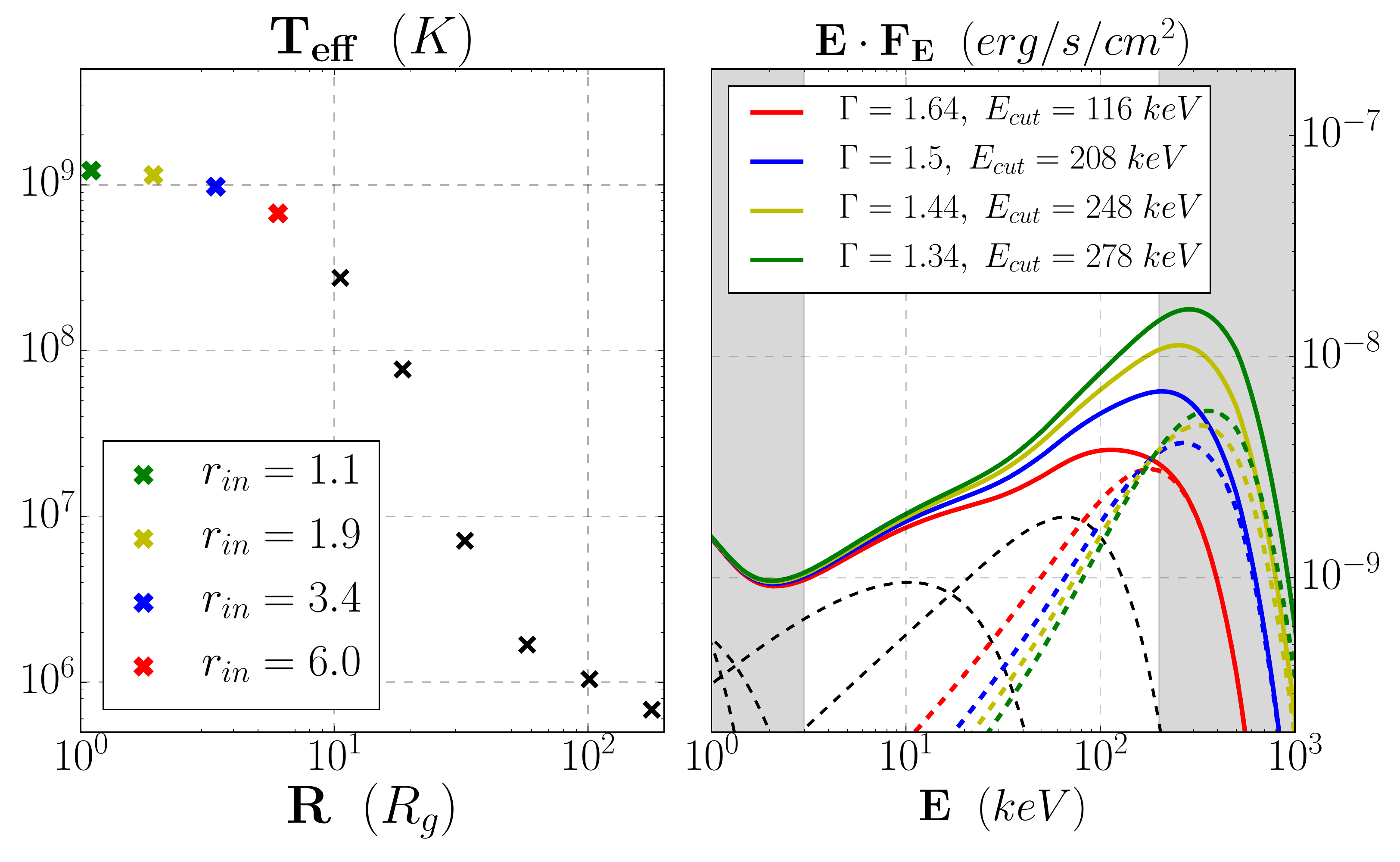}
  \caption{Slim, optically thick JED configuration with $\dot{m}_{in} = 2$ at $r_{in}=1$ and ($\mu=0.1, \xi=0.1,m_s=1.5, b=0.3$). Left: Effective temperature as a function of the radius. 
  Right: Spectrum emitted at each disk radius $r$ (dashed lines, same color code as in the left panel). The solid lines are the overall disk spectra, i.e., the sum of the contributions from each disk annulus down to the inner radius $r_{in}$. The inset shows the corresponding photon index $\Gamma$ and high-energy cutoff $E_{cut}$, obtained with a fit $E^{2-\Gamma} e^{- (E/E_{cut})}$.}
   \label{fig:EffectOfRi_OnSlim}
\end{figure}

It is worthwhile, however, to assess the effect of $r_{in}$ on slim JED solutions better, as shown in Fig.~\ref{fig:EffectOfRi_OnSlim}. We computed the thermal balance of a JED established from $r_{in}=1$ with $\dot{m}_{in} = 2$ and the dynamical parameters ($\mu=0.1,~ \xi=0.1,~ m_s=1.5, \text{and}~ b=0.3$). For these values, the inner disk regions are optically thick and geometrically thick (slim), with an effective temperature displayed on the left side of the figure. In order to test only the variation of the last innermost orbit on the overall disk spectrum (same $\dot m$), we show in the right panel of Fig.~\ref{fig:EffectOfRi_OnSlim} the contributions of each disk annulus (dashed lines). The solid lines display the overall disk spectrum, that is, their sum up to the last radius $r_{in}$. Going from $r_{in}=6$ (red) to $r_{in}=1.1$ adds more and more high-energetic slim components. This affects not only the total emitted power, but also the shape of the spectrum, that is, the photon index $\Gamma$ and high-energy cutoff $E_{cut}$. Using the same spectral fit procedure, we see that the spectrum changes from $(\Gamma=1.64, E_{cut} \simeq 100~\text{keV})$ to  $(\Gamma=1.34, E_{cut} \simeq 300~\text{keV})$. Quite obviously, these numbers must be taken with great caution, as they were obtained within a Newtonian approach, but they nevertheless reveal the tendency of how the black hole spin would affect the high-energy cutoff of XrBs during their high-luminosity hard states.

\subsection{Accretion Mach number $m_s$}

\begin{figure}[h!]
  \centering
  \includegraphics[width=\columnwidth]{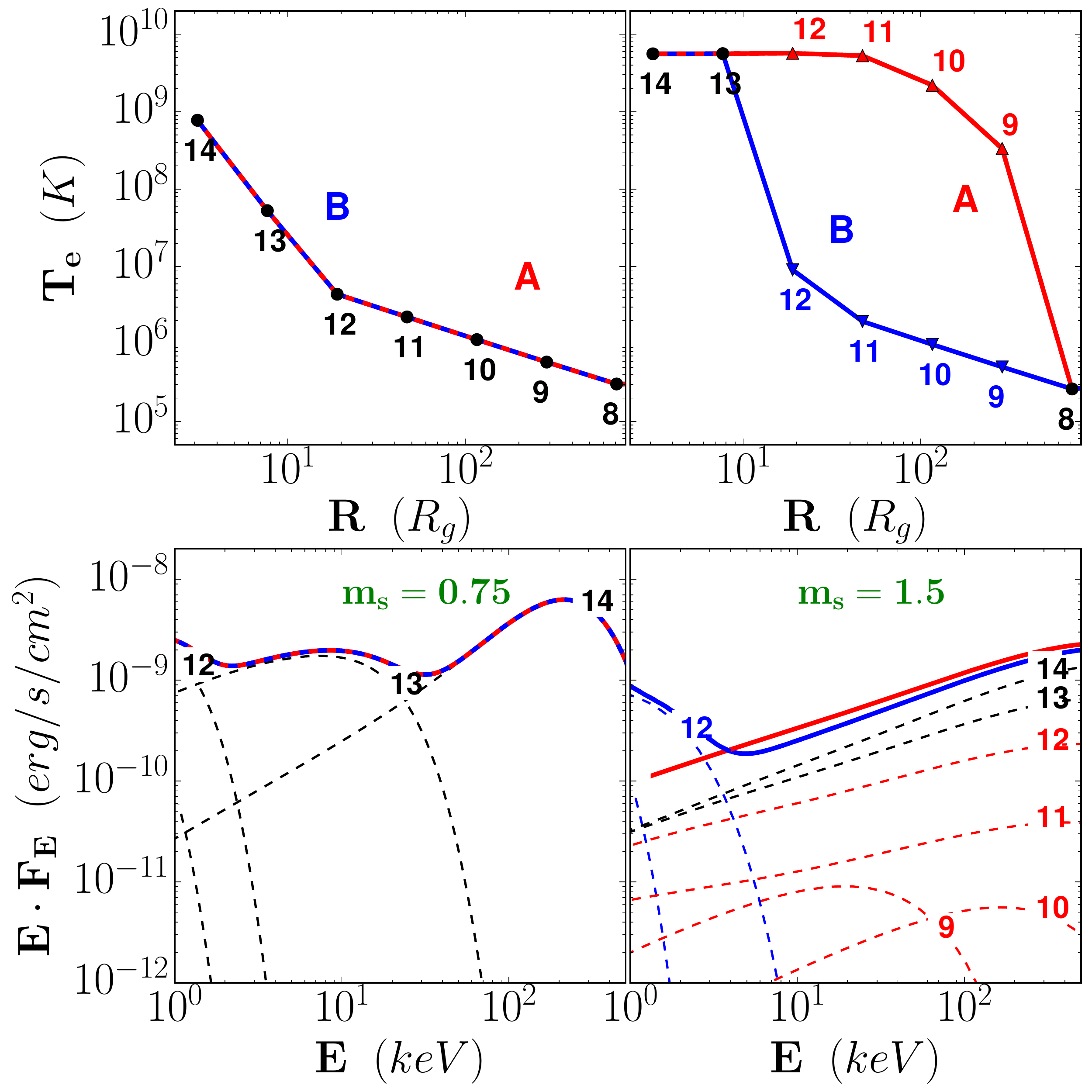}
  \caption{Effect of the accretion Mach number $m_s$ on a JED configuration fed with $\dot{m}_{in} = 0.5$ at $r_{in}=3$, $m_s= 0.75$ (left) and $
  m_s= 1.5$ (right). This figure is similar to Fig.~\ref{fig:EffectOfB}.}
  \label{fig:EffectOfMs}
\end{figure}

The Mach number $m_s= - u_R/c_s$ measures the strength of the torque acting upon the disk plasma. Thus, for a given (local) accretion rate, the disk column density is $\Sigma= - \dot M(r)/ 4 \pi R u_R$: the smaller $\Sigma,$ the larger $m_s$. It has been shown that in a JED, accretion is trans-sonic with $m_s \gtrsim 1$ \citep{Ferreira97}. We present in Fig.~\ref{fig:EffectOfMs} two different JED calculations fed with $\dot{m}_{in} = 0.5$ at $r_{in}=3$, obtained for  $m_s = 0.75$ (left) and $m_s= 1.5$ (right). 

Going from $m_s=0.75$ to $m_s=1.5$ leads to a decrease by a factor 2 in the disk density and changes the JED solutions in two ways. First, it allows the possibility of geometrically thick, optically thin solutions for $m_s=1.5$ (hence two distinct configurations A and B), whereas $m_s=0.75$ leads only to the optically thick solution (configurations A and B are identical in this example).
Second, the optically thick disk solution also appears to be warmer. The reason is that the slim solution appears much sooner. In the $m_s=0.75$ solution, it occurs only around point 14 ($r\simeq 6$), whereas it appears already around point 12 ($r\simeq 12$) for $m_s=1.5$. Below this radius, advection takes over and the optically thick disk becomes hotter. Advection starts to play a role sooner because the disk is more tenuous and hence has a less efficient Coulomb thermalization that leads to $T_i\simeq 24 T_e$ here (whereas $T_i=T_e$ for $m_s=0.75$).

This result also depends on the value of the disk accretion rate. For lower values, for instance, no slim-disk solution would be possible and varying $m_s$ would have no drastic effect. This study shows, however, that the precise relation between the density and the dominant torque may have a tremendous effect on the overall disk spectrum (see section \ref{sec:bestparams}).

\subsection{Ejection efficiency $\xi$}
\label{sec:EffectOfXi}

\begin{figure}[h!]
  \centering
  \includegraphics[width=\columnwidth]{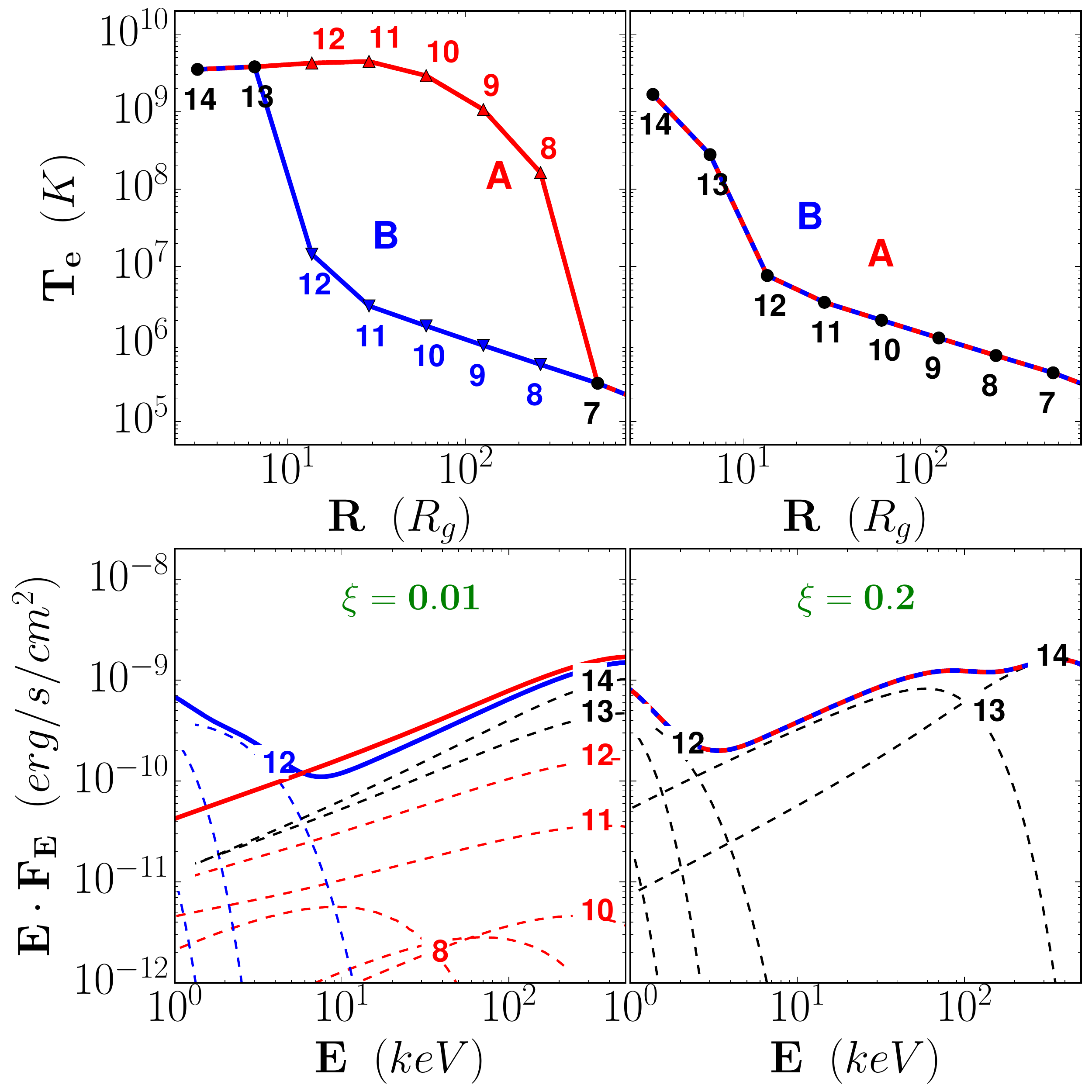}
  \caption{Effect of the disk ejection efficiency $\xi$ on a JED configuration fed with $\dot{m}_{in} = 0.6$ at $r_{in}=3$, $\xi= 0.01$ (left) and $
  \xi = 0.2$ (right). This figure is similar to Fig.~\ref{fig:EffectOfB}.}
  \label{fig:EffectOfXi}
\end{figure}

In a realistic accretion disk, a complex function $\dot m(r)$ might be expected that describes the local accumulation or decrease of the column density $\Sigma$ as well as some mass loss, whose efficiency could vary from one disk region to another. In our steady-state simplified JED picture, $\dot m \propto r^\xi$ describes a disk giving rise to mass loss with a local ejection efficiency $\xi$ that is constant everywhere. It has been shown that this ejection efficiency is both a function of the disk magnetization $\mu$ \citep{Ferreira97} and thermal aspects at the disk surface layers \citep{Casse00b, Murphy10, Tzeferacos13}. For the sake of simplicity, however, we use $\xi$ and $\mu$ as independent parameters. 
 
Figure~\ref{fig:EffectOfXi} shows two JED calculations fed with $\dot{m}_{in} = 0.6$ at $r_{in}=3$, one for a typical cold jet with $\xi = 0.01$ (left) and the other for a more massive jet with $\xi=0.2$. We recall that the accretion rate in the outer disk regions increases with $\xi$
despite the same $\dot m_{in}$. For instance, at radius $r\simeq 320$ (point 8), the heavy jet solution (right) is fed with $\dot m = \dot m_{in} \left( \sfrac{320}{3} \right) ^{0.2}= 2.54 \dot m_{in}$, while for the more tenuous jet $\dot m = \dot m_{in} \left( \sfrac{320}{3} \right)^{0.01}= 1.05 \dot m_{in}$, which
is a difference of a factor $2.4$ . This explains the different behavior of the two JED calculations. For $\xi=0.01$, the inner regions below point $7$ exhibit the two thermally stable solutions (hence the two configurations A and B), while the heavy mass-loss solution only displays the geometrically thin solution: the disk is too dense to allow a geometrically thick solution with $T_e \neq T_i$ to exist. 

In the inner regions, below point 12, the optically thick solution gradually becomes a slim JED in both cases. However, the less dense disk (left) leads to a stronger influence of advection and thereby to a hotter plasma. As a consequence, the configuration B spectrum reaches a higher energy for $\xi=0.01$ than for $\xi=0.2$. We conclude that the disk ejection efficiency has an effect on the global disk spectrum; this might be tested observationally.

\section{Thermal hysteresis cycle}
\label{sec:Hysteresis}

The situation of two thermally stable and one intermediate unstable solution
within an interval of radii and for a range of
disk accretion rates is reminiscent of the so-called S-curve in accretion disk theory \citep[see, e.g.,][]{Frank92}. Such a property naturally leads to a hysteresis cycle when the disk accretion rate evolves in time. The question is whether  this JED thermal hysteresis can account for typical XrB q-shaped cycles.

In this section, we perform a very simple study where we assume that the dynamical JED parameters are frozen during the entire evolution and are thus the same in configurations A and B. Being able to vary these parameter would certainly allow us to perform a parametric cycle, but this is beyond the scope of this paper. We therefore choose a JED configuration settled from $r_{in}=1$ to $r_{out}=10^3$ and the dynamical parameters  ($\mu= 0.1,~ \xi= 0.01,~ m_s=1.5,~ b=0.7$).

\begin{figure}[h!]
  \centering
  \includegraphics[width=\columnwidth]{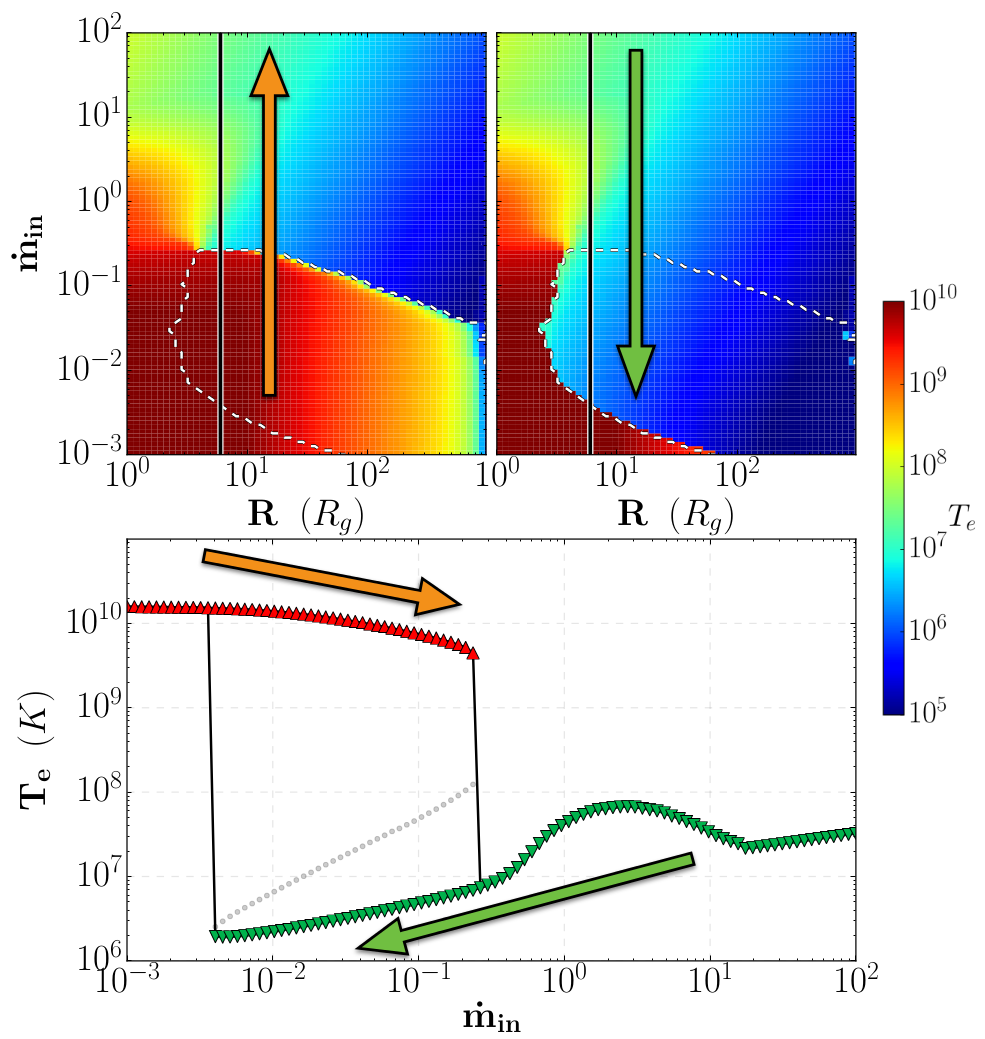}
  \caption{Top: JED ($\dot{m}_{in}-r$) solution plane for ($\mu= 0.1,~ \xi= 0.01,~ m_s=1.5,~ b=0.7$). The background color shows the electron temperature (see colorbar in Kelvin on the right side) for the hottest solution (left) and the coolest solution (right). The dashed white line outlines the zone where these two thermally stable solutions coexist. The slice at $r=5$ (vertical solid line) shows the two solutions as function of $\dot{m}_{in}$, shown in the bottom panel (a version of the S-curve). The red upper triangles show the thick/hot disk solution, and green lower-triangles
represent the thin/slim cooler disk. The third, thermally unstable solution is shown as gray dots. The arrows describe a possible hysteresis cycle (see text).}
  \label{fig:Hysteresis}
\end{figure}

Figure~\ref{fig:Hysteresis} shows all computed thermal JED solutions for a range of $\dot m_{in}$  from $10^{-3}$ to $10^2$ \citep[see also Fig. 2 in][]{Petrucci10}. The top left ($\dot{m}_{in}-r$) panel shows in background color the electron temperature of the hottest solution found at each radius, and the top right panel
shows the coolest solution we found. The zone where these two images differ is outlined by a dashed white line: within this zone, two possible solutions can coexist, one cold and optically thick and the other hot and optically thin. For the dynamical parameters used here, this zone is located somewhere between $10^{-3} < \dot{m}_{in} < 3 \times 10^{-1}$ and $3 < r < 2 \times 10^{2}$. We stress that the vertical axis is the accretion rate $\dot m_{in}$ and not the accretion rate measured at a given radius. A steady-state JED configuration is thus a horizontal slice with fixed $\dot{m}_{in}$ and $\dot{m} (r)$ variable. For instance, for $\dot m_{in}=0.1$, the outer disk is in a cool state down to $r\sim 2 \times 10^{2}$ where the solution forks, designing the two familiar A (hot) and B (cold) configurations.

In the bottom panel of figure~\ref{fig:Hysteresis}, we show a vertical slice of the two top panels at $r = 5$. This figure shows the electron temperature for all possible accretion rates, ranging from $\dot{m}_{in} = 10^{-3}$ to 100. Assuming the disk remains on the same branch, the hysteresis cycle can proceed as follows: starting with a low accretion rate and increasing it until  $\dot m_{in}=10$,  for instance, leads in this outbursting phase to the following behavior, sketched with the orange arrow.  The annulus first follows a succession of nearby hot solutions (red triangles) until $\dot m_{in,h}\simeq 0.3$, where the disk no longer finds such a solution and bifurcates toward the optically thick cold solution (green triangles). The rise in accretion rate is now shown along the lower solution track. When the system undergoes a decline, with $\dot m_{in}$ decreasing toward its initial value, the evolution of the annulus follows a different path sketched with the green arrow. The annulus remains cool and geometrically thin until $\dot m_{in,l} \simeq 0.004$, where the solution bifurcates toward the hot and geometrically thick upper branch.

This scenario is quite appealing. However, accretion of material is triggered from the outer disk regions. As a consequence, the incoming plasma depends of the outer state and we therefore need to consider the disk as a whole. Within our simplified steady-state picture, this implies considering the change in disk configurations (and not only a local radius) as the accretion rate evolves in time. This is sketched with the orange (for the rise in $\dot m_{in}$) and green (decline) arrows in the top panels of Fig.~\ref{fig:Hysteresis}. Starting again with $\dot m_{in}= 0.001$, the JED is in configuration A, that is, optically thin and geometrically thick up to a transition radius $r_t \simeq 1000$, beyond which it is optically thick and geometrically thin. As the disk accretion rate increases (orange arrow), the luminosity increases and $r_t$ gradually decreases. A major spectral change occurs when $\dot m_{in}$ reaches a critical high value $\dot m_{in,h}$ (close to $0.1-0.2$) where the optically thin solution disappears (configuration B everywhere). In the decline phase (green arrow), there will be an inside-out reconstruction of the hot and geometrically thick disk (configuration A), starting at a critical low value $\dot m_{in,l}$ (presumably close to 0.01).   

Although this simple scenario reproduces some of the qualitative properties of the hysteresis cycle in X-ray binaries, it fails to reproduce some more precise observations. 
\begin{enumerate}

\item When we identify configuration A with the hard states and configuration B with the soft states, we would find a hard-to-soft transition for $\dot m_{in,h} \sim 0.2$ and a soft-to-hard transition for $\dot m_{in,l} \sim 0.01$. In terms of luminosity, calculated by integration of the global disk spectrum, this would correspond to roughly  $L_{H-S} = 0.02\,  L_{Edd}$ and  $L_{S-H} = 0.001\, L_{Edd}$ , respectively. These values depend on the parameter set chosen (see Fig.~\ref{fig:Gamma_Ecut}), but the hysteresis area seems to remain too low in luminosity in any case. Both values are lower (typically by one order of magnitude) than common observed transition-state luminosities \citep[e.g.,][]{Dunn10}.

\item This scenario assumes constant dynamical JED parameters, the evolution being due to the variation of the control parameter $\dot m_{in}$ alone. Producing both soft and hard state spectra with the same parameter set during the round-trip is much more difficult than it appears, however. 
Most of the time, the spectral shape of the solution reproduces hard states very well but soft states only very rarely. The reason
is that although the thinnest possible solution (top right panel of Fig.~\ref{fig:Hysteresis}) is highly dominated by thin-disk solutions, the inner part of this configuration B is mostly in the thick/slim disk state of its hysteresis area. The resulting spectral shape is dominated by these inner regions of the disk and the spectrum is quite different from a soft state (see, e.g., the configuration B spectrum shown in Fig.~\ref{fig:EffectOfMs}).

\item Finally, observationally, the hard-to-soft state transition is accompanied by a huge decrease in radio luminosity, and no jet detection has ever been reported during soft states. This is commonly interpreted as a quenching of the jet. In this case, JED solutions are not expected by definition, and soft states would be reproduced by pure SAD. However, it has recently been argued that a jet could always be present, but would simply not radiate in the soft states for the acceleration mechanisms in the jet would be inefficient \citep[][and the "dark jet" model]{Drappeau17}. In this case, JED solutions would still be relevant.   
  
\end{enumerate}

\begin{figure*}[ht]
  \centering
  \includegraphics[width=\textwidth]{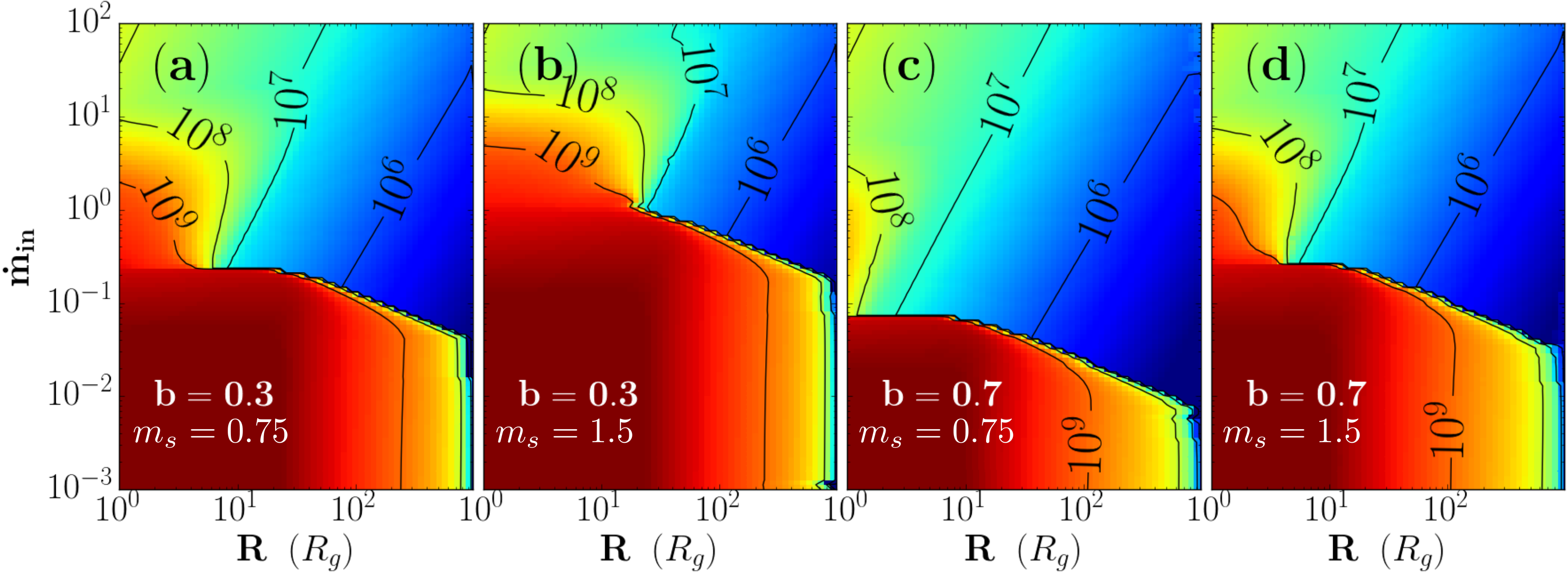}
  \caption{JED $\dot{m}_{in}-r$ hard-state solution plane for $r_{in}=1$ and $\mu= 0.1, \xi= 0.01$. The color background shows the hottest solution  and solid lines are isothermal contours (in Kelvin). From left to right: (a) $m_s=0.75,~ b=0.3$; (b) $m_s=1.5,~ b=0.3$; (c) $m_s=0.75,~ b=0.7,$ and (d) $m_s=1.5,~ b=0.7$. The colorscale in this figure is the same as in Fig.~\ref{fig:Hysteresis}.}
  \label{fig:ThickThinSlim}
\end{figure*}

\begin{figure*}[ht] 
   \centering
   \includegraphics[width=\textwidth]{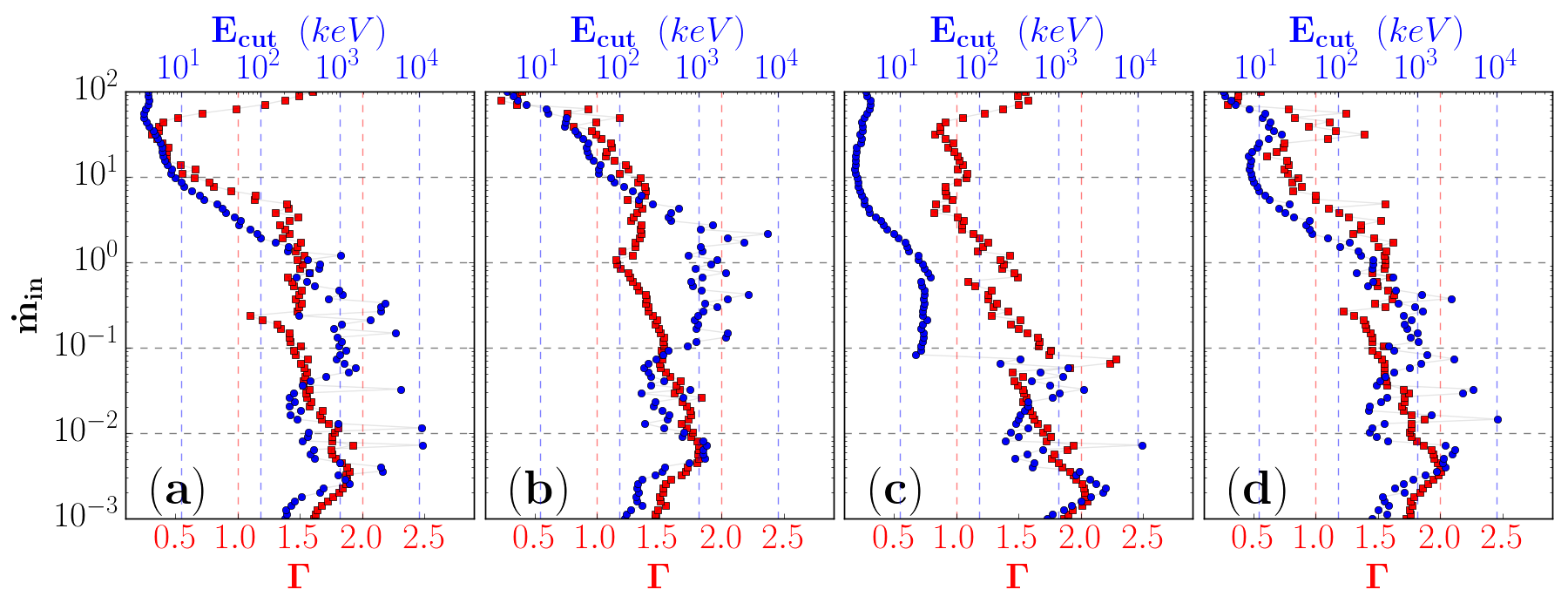}
      \caption{Photon index $\Gamma$ (red) and high-energy cutoff $E_{cut}$ (blue) as function of $\dot m_{in}$, resulting from the spectral fit of each JED configurations (a), (b), (c), and
(d) displayed in Fig~\ref{fig:ThickThinSlim}. No error bars on fits are displayed here for presentation purposes; errors on $\Gamma$ are on the order of $0.1-0.2$, and errors on $E_{cut}$ are up to few hundreds when $E_{cut} > 100$~keV and only few percent when $E_{cut} \lesssim 100$~keV.}
         \label{fig:Gamma_Ecut}
\end{figure*}  

These three independent arguments strongly argue for a real dynamical process that would link the disk spectral state with jet production. 
%
%
Combining these elements and following our initial suggestion \citep{Ferreira06, Petrucci08}, we hereafter propose that the rise and fall in $\dot m_{in}$ is accompanied by an evolution of the disk magnetization $\mu(r)$ that, in turn, triggers the transition from a SAD to a JED and vice versa. Clearly, this rationale leads to the simple picture where soft states would be described by internal SAD configurations whereas hard states would require internal JED configurations. A detailed investigation of SAD-JED hybrid configurations will be addressed in a forthcoming companion paper.

\section{JEDs in hard states}
\label{sec:bestparams}

In this section, we study the hard states produced by a JED along the rising phase of XrBs in greater detail. In particular, we wish to assess the most favorable parameters (assumed constant during the outburst) that allow us to describe hard states up to high luminosities.

Hard states of XrBs are characterized by a power-law spectrum of index $\Gamma$ ranging from 1.3 to 1.8, and a hardly detectable blackbody component. During the rising phase, the disk luminosity varies from $L_{tot} \sim 10^{-3} L_{Edd}$ to $L_{tot} \sim 0.1\,  L_{Edd}$ or even higher, maintaining a power-law dominated spectrum although $\Gamma$ varies \citep{Dunn10}.
If the inner parts of accretion disks accrete under the JED mode, then one must be able to reproduce the above behavior within a JED configuration. According to the parameter study done before, the parameters that are important (producing the strongest modifications in the spectra) are the accretion Mach number $m_s$ and the jet power fraction $b$. The others, $\mu$ and $\xi$, have a weaker
effect on the overall disk spectrum, and their value can also be constrained by dynamical arguments. We hereafter use the conservative values $\mu=0.1$ and $\xi=0.01$. We also decide to fix $r_{in}=1$ in order to facilitate our representation in the $\dot m_{in}-r$ solution plane. If, for instance, it were to be decided that $r_{in}$ should be 6, solutions below $r = 6$ can be disregarded\footnote{This can be done here since no inside-out effects are present in this paradigm.}. 

We can now create a $\dot m_{in}-r$ JED solution plane designed to completely reproduce the hard states, chosing a wide range in accretion rate that completely includes the hard-state luminosity range. This implies that only configuration A is displayed, where the hot optically thin solution exists. This is shown in Fig.~\ref{fig:ThickThinSlim} for four parameter sets (from left to right): (a) $m_s=0.75,~ b=0.3$; (b) $m_s=1.5,~ b=0.3$; (c) $m_s=0.75,~ b=0.7,$ and (d) $m_s=1.5,~ b=0.7)$. For each parameter set, we also display in Fig.~\ref{fig:Gamma_Ecut} the photon index $\Gamma$ and cutoff $E_{cut}$ values obtained through a simple fit ($E F_E = E^{2-\Gamma} \cdot e^{-E/E_{cut}}$) of the total spectrum at each $\dot m_{in}$. 

At very low accretion rates $\dot{m}_{in} \sim 10^{-3}-10^{-1}$ (bottom of Fig.~\ref{fig:ThickThinSlim}), all JED configurations are cold/optically thick in the outer parts ($r>1000$) and hot/optically thin in the inner parts of the disk. The spectrum is of course dominated by the innermost regions where $T_e \gtrsim 10^9$ K and $\tau_{tot} \ll 1$. Figure~\ref{fig:Gamma_Ecut} allows us to be more specific. The JED parameter set (c) produces harder spectra that become soft\footnote{$E_{cut} < 10-20$ keV indicates that the spectrum is much closer to a disk blackbody than a power law. Law values of $\Gamma$ in these cases are a consequence of the basic spectral fitting procedure used.} much sooner (around $\dot m_{in} \sim 0.1$) than the other three sets. This set can therefore be ruled out to explain high-luminosity hard states. The other JED configurations produce power-law spectra with a high-energy cutoff higher than $100-200$~keV as in Fig.~\ref{fig:Exple_sed} (red spectrum), which is compatible with the observations. 

At high accretion rates, for example, $\dot{m}_{in}$ between 1 and 10 (top of the diagrams), JED configurations (a), (b), and (d) are in the slim state. These solutions have a typical inner temperature $T_e \simeq 10^8-10^9$~K and are optically thick, although the spectrum mimics that of a hard state (see section \ref{sec:Slim}). The spectra of sets (a) and (d) are too hard ($\Gamma$ reaches unity) and the energy cutoffs are too low (around 10 keV), which is not fully consistent with the observations of high-luminosity hard states. In contrast, set (b) maintains a power-law spectrum with a typical spectral index $\Gamma \sim 1.5$ and high-energy cutoff $E_{cut}$ higher than but on the order of 100~ keV. 

The situation becomes more complex at intermediate accretion rates around $\dot{m}_{in} \sim 0.1-1$ (middle of the diagrams), as the outcome now strongly depends on the parameters $(m_s, b)$.  For a given $\dot{m}_{in}$, some configurations will still be in the optically thin regime (e.g., $a, b, d$), while others ($c$) have switched to the optically thick, geometrically thin cold ($T_e = 10^6-10^7$~K) solution with a multi-blackbody spectrum as in Fig.~\ref{fig:Exple_sed} (blue spectrum).  

XrBs do not show any evidence of such a strong multi-blackbody spectrum while ascending the vertical hard-state branch. Regardless
of the dynamical and spectral modifications they undergo, the spectrum remains dominated by a power law. Evolving from low to high $\dot{m}_{in}$, a JED configuration needs to shift continuously from the hot/optically thin solution to the slim solution at the innermost radii. This therefore favors dynamical accretion modes that can provide the highest accretion speed but still
retain a significant fraction of the accretion energy released as radiation within the disk. This simple general trend seen in XrBs clearly dismisses parameter sets (a), (c) and (d), while the parameter set (b) with $m_s=1.5, b=0.3$ appears to fulfill all observational requirements.

\begin{figure}[h] 
   \centering
   \includegraphics[width=\columnwidth]{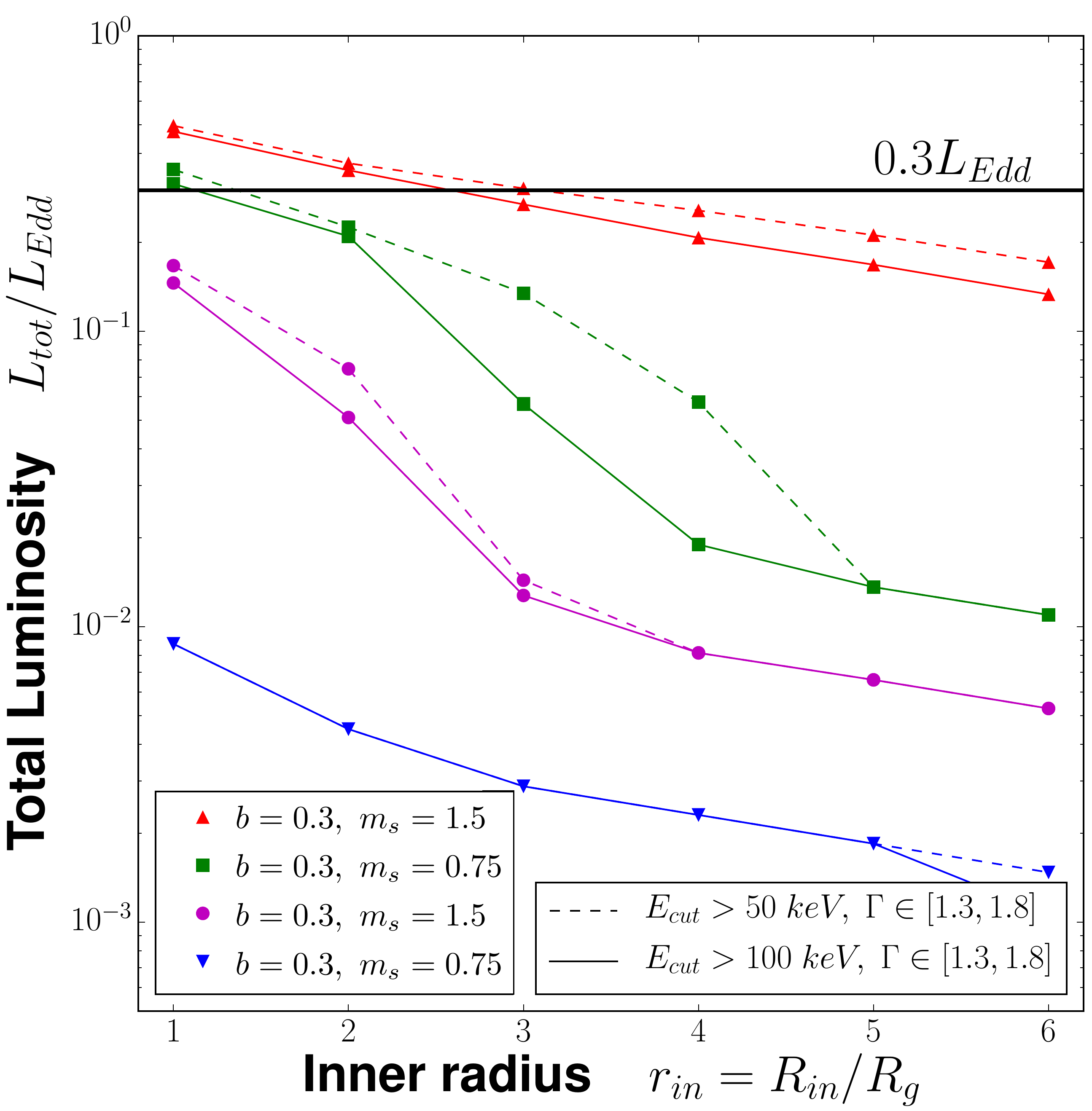}
      \caption{Most luminous possible hard-state for our four different parameter sets [$b=0.3-0.7$ and $m_s=0.75-1.5$] as a function of internal radius in different colors and markers.}   
         \label{fig:Lum_rin}
\end{figure}  

This is correct, however, in particular at high luminosities ($\dot{m}_{in}$ higher than a few), only if $r_{in}$ is small enough. Within our simplified Newtonian approach, we find that $r_{in} <3$ is necessary for the parameter set (b) to be able to build high-luminosity hard states, $L_{tot} > 30 \% L_{Edd}$ with $\Gamma \in [1.3, ~ 1.8]$ and $E_{cut-off} > 50$ or $100$ keV. This is displayed in Fig.~\ref{fig:Lum_rin}, where we present the most luminous possible hard state in each of the configurations. This both confirms that the parameter set chosen [$b=0.3$ and $m_s=1.5$] is the best of the four we presented, and that the disk must reach very small radii $r_{in} \simeq 2-3 ~ R_g$ to be able to produce spectra with higher luminosity. This inner radius corresponds to a black hole spinning at $0.78 < a_{th} < 0.94$, which is consistent with some observational studies from GX 339-4 that found $a_{obs} \sim 0.9-0.94$ \citep[see, e.g.,][]{Miller08, Reis08, Kolehmainen10, Plant15}.

\section{Concluding remarks}

We have shown that a jet-emitting disk, described by a single parameter set, is able to qualitatively reproduce the spectra of the archetypical object GX 339-4 during its hard state, from low to high luminosities with the evolution of the accretion rate. This is very encouraging and deserves further developments. There are, however, several simplifications and caveats that require some caution and call for improvements in the future.

\subsection{Caveats}

For the sake of simplicity, we assumed a Keplerian rotation law down to $r_{in}$. While in a SAD the deviation from this law is on the order $\varepsilon^2$, in a JED, it is on the order of $\mu \varepsilon$ (Ferreira \& Pelletier 1995). We obtained JED solutions reaching a disk aspect ratio $\varepsilon \simeq 0.3$, which may indeed provoke a non-negligible deviation. In
addition, we assumed a Newtonian accretion disk while allowing the disk inner radius to reach values of about unity. Our calculations
therefore need to be reevaluated by solving the radial disk equilibrium in a pseudo-Newtonian potential, for instance (Paczynski \& Wiita 1980). In a consistent way, relativistic effects as well in the treatment of the radiation (beaming, ray tracing) and inclination effects need to be included. This is clearly beyond the scope of this paper, but deserves further investigations. However, none of these improvements are likely to modify the main conclusions we presented. \\

We assumed for simplicity that the local disk scale height $H(R)$ is correctly described by a quasi-hydrostatic prescription (see eq.~\ref{eq:vert}). However, this is not quite the case in a JED. As discussed in \citet{Ferreira95}, the mean-field magnetic configuration produces a vertical pinching of the disk through both its bending ($B_r$ component) and shear ($B_\phi$ component). It has been found that the correct JED scale height is smaller than the hydrostatic value by a factor of approximately between 0.3 and 0.95, depending on the details of the vertical structure
of the disk (such as the value of the disk ejection efficiency $\xi$). As a consequence, our analysis is not entirely consistent. However, taking this magnetic pinching into account
would also require addressing the link between the thermal state of the disk and the jet mass loss, for instance, which are beyond
both our current knowledge and the aim of this paper. Moreover, the turbulent magnetic field, which has been neglected so far in our current mean-field approach, may also lead to an additional pressure term that puffs out the disk. The final outcome of these two competing effects is quite difficult to envision. The same difficulty also applies to the computed spectra from the optically thin disk regions because of the nonlinearity in $\epsilon$ introduced by the inverse Compton effect. Removing this hydrostatic approximation is a step that should probably be made in a further development of the model. \\

The parameters used to describe the dynamical JED accretion mode have been assumed to be constant. This is a twofold simplifying assumption. 

\noindent The first simplification concerns their evolution in time. Each spectral state that we compute assumes a steady-state accretion disk, where any radial evolution of the disk accretion rate is due to mass loss and not to a transient modification of the disk density $\Sigma$. We are aware that this may be a strong limitation of our approach and that a fully time-dependent treatment, such as in the disk instability model or DIM \citep[][and references therein]{Hameury17}, should be made. However, the DIM computes only SAD configurations with simplified spectra and is thereby unable to address the points discussed in this paper. Notwithstanding these difficulties, our approach provides results that will, hopefully, lead to the development of a more complete version of the DIM. 

\noindent The second simplification concerns the radial distributions of the dynamical parameters in a steady-state disk. The constant values chosen for the JED configuration ($\mu= 0.1,~ \xi= 0.01,~ m_s=1.5,~ b=0.3$) should not be considered too strictly. However, they lie within the parameter space that has been explored with self-similar accretion-ejection structures \citep[see][]{Ferreira97, Casse00a, Casse00b} and allow reproducing GX 339-4 hard states, which will be demonstrated in a forthcoming paper. This is very exciting for such a simplified model. \\

The low value chosen for the ejection efficiency $\xi$ also raises some questions. For simplicity, we used the value $\xi=0.01,$ which is typical for cold, magnetically driven jets (Ferreira 97). The ejection efficiency might be increased quite easily
when thermal effects come into play at the disk surface \citep{Casse00b, Murphy10, Tzeferacos13, Bethune17}. Of all JED parameters, $\xi$ might therefore vary the most with the radius. Evidence of winds in XrBs might even be explained if a JED with $\xi>0.1$ or larger is settled at the outermost disk radii \citep[see][and references therein for more details]{Fukumura10, Chakravorty16}. On the other hand, $\xi$ is also related to the maximum asymptotic Lorentz factor that is achievable in a magnetized jet, and we might therefore try to constrain it from observations \citep[see][for the Cygnus X-1 case]{Petrucci10}. \\

Finally, although our treatment of the optically thin radiative processes is quite successful, it remains simplified when compared to observed spectra. 

\noindent First, the \textsc{BELM} code assumes a one-zone model of constant temperature, density, and magnetic field that we identified as the midplane disk values at a given radius. However, a disk is vertically stratified and a full radiative transfer should probably be performed, as shown, for example, in \citet{Ross96} or \citet{Davis06}.

\noindent Second, no reflection components have been included in our model, but all XrB spectra show evidence of such components. As a consequence, our theoretical spectra cannot match the observational data perfectly. Fortunately, these reflection components neither represent the major portion of the detected flux nor the most important features, which is the reason why we disregarded them \citep[see][and references therein]{Plant14}. It is currently believed that these components are due to a reflection of a hard X-ray spectrum on an ionized optically thick medium located in the innermost disk regions. In our study, the inner parts of
the disk would only be optically thick in the high-luminosity, transitional hard states that would correspond to our slim JED solutions ($\tau_{tot} \sim 1-10$).  

As argued for the simplifying assumptions made on the disk dynamics, however, we doubt that our main conclusions would be strongly affected by taking into account a more realistic treatment of radiation.

\subsection{Summary}

We here extended the paradigm proposed in paper I \citep{Ferreira06} by computing the thermal balance in an accretion disk settled from an innermost radius $r_{in}$ to $r_{out}$ and fed with an accretion rate $\dot m_{in}= \dot m(r_{in})$. Within our paradigm, the rise and fall in disk accretion rate $\dot m_{in}$ is accompanied by a modification of the local disk magnetization $\mu= B_z^2/\mu_o P_{tot}$ that in turn determines the dominant torque and allows accretion. If $\mu$ is greater than a critical value $\mu_c \sim 0.1,$ the disk launches jets \citep[see paper I and][]{Petrucci08} that vertically carry away the disk angular momentum (JED mode), while for a weaker magnetization, the disk is in the SAD mode where turbulence transfers the disk angular momentum outward in the radial direction. We focused on the thermal structure and emitted spectra from a JED.

We developed a two-temperature plasma code that is able to address optically thin/thick transitions, radiation, and gas-supported regimes and that can compute the emitted spectrum from a steady-state JED in a consistent way. The radiative processes taken into account are bremsstrahlung and synchrotron emission and their local Comptonization through the \textsc{BELM} code \citep{Belmont08, Belmont09}. Our procedure incorporates energy advection, starting from $r_{out}$ and progressing outside-in. Advection carries to the smaller radii the outer radius evolution. Our code is therefore designed to address the disk inner regions, whichare responsible for the spectral domain from optical to X-rays.\\ 

(1) We showed for the first time spectra of jet-emitting disks that were computed self-consistently, along with the resolution of the thermal balance of the disk. We recovered the results obtained by \citet{Petrucci10} with their one-temperature JED calculations, that is, the possibility of two thermally stable solutions at a given radius, for a wide range in disk accretion rates. This opens up the possibility of a hysteresis cycle for JEDs. Although quite compelling, we showed that such a cycle would hardly be compatible with those observed in XrBs as the transitional (i.e., hard-to-soft and soft-to-hard) luminosities would be too low.
In addition, jets would be always dynamically present, and an explanation would be needed why they would not shine during the soft states \citep[but see an example of such a process in][]{Drappeau17}. The simplest (and commonly accepted) understanding of XrB cycles is that the hard-to-soft spectral transition leads to (or is the signature of) a dynamical quenching of the jet, while the soft-to-hard transition corresponds to the start of jet production. Within our paradigm, this translates into a dominant JED mode while in the hard state and a dominant SAD mode while in the soft state.\\

(2) We reported hot, optically thick SLIM solutions for the JED mode at high $\dot m_{in}$ that produce power-law spectra similar to those usually attributed to an optically thin emission. We show that for a reasonable set of parameters, the dynamics imposed by magnetically launched jets gives rise to spectral signatures consistent with the hard states. To be precise, the evolution of both the photon index $\Gamma$ and high-energy cutoff $E_{cut}$ appears to be generally consistent with observations from low to almost near Eddington luminosities ($\Gamma \in [1.3,~1.8]$ and $E_{cut} \simeq 50$ keV appearing at high luminosities only).\\  

(3) Although our calculations were made using a Newtonian approximation, we allowed the innermost disk radius $r_{in}$ to vary down to $r_{in} =1$ for a few illustrations, and $r_{in} \sim 2-3$ for most of the simulations. Only $r_{in}< 3$ allows both a smooth spectral transition from optically thin-slim JED solutions and enough energy to produce high-luminosity hard states. If this trend is confirmed by relativistic calculations (a work to be done), then it would provide a means to constrain the black hole spin.\\

Finally, explaining large XrB outburst cycles requires the disk to switch from one accretion mode to another (as first proposed in paper I). Since such a mode transition cannot occur simultaneously across the whole disk \citep{Petrucci08}, the disk must be described at any given time by some hybrid JED-SAD configuration. This will be explored in a forthcoming paper.

\begin{acknowledgements}
      We are grateful to the anonymous referee for their careful reading of the paper. The authors acknowledge funding support from French Research National Agency (CHAOS project ANR-12-BS05-0009 http://www.chaos-project.fr), Centre National de l'Enseignement Supérieur (CNES) and Programme National des Hautes Énergies (PNHE) in France. SC is supported by the SERB National Postdoctoral Fellowship (File No. PDF/2017/000841).
\end{acknowledgements}

\bibliographystyle{aa} 
\bibliography{ADSbibnew.bib}

\appendix 
\appendix 

\section{Numerical review}
\label{sec:Method}

\subsection{Global method of resolution}

In order to obtain the thermal state of the disk, the full set of equations described in Sect.~2 needs to be solved. This set is reduced to a nonlinear system of equations expressed with three independent variables: the electron temperature $T_e$, the ion temperature $T_i$ , and the disk aspect ratio $\varepsilon=H(R)/R$ ($P_{rad}$ is computed using Eq.~\ref{eq:rad}). Some parameters are object dependent (mass $m$, innermost and outermost disk radii $r_{in}$ and $r_{out}$), while others are related to the dominant disk accretion mode under consideration (disk magnetization $\mu$, accretion Mach number $m_s$, ejection efficiency $\xi$ , and jet power fraction $b$ for the JED). The disk accretion rate $\dot m_{in}$ is used as a varying control parameter that
allows scanning the various spectral states of the disk. 

A cylindrical symmetry is assumed and the disk is discretized in annuli of volume $2 \pi R \times dR \times 2H$. We choose a logarithmic grid with $dR / R = \text{Cst} \sim 0.1$, which is a good compromise between calculation time, error rate, and precision, while being physically and numerically correct (see following section \ref{sec:EffectofNdr}). We use an optically thick-thin bridge formula for the radiative cooling rate $q_{rad}$ and therefore need to evaluate both optically thick $q_{thick}$ and optically thin $q_{thin}$ cooling rates. While the former is analytical, the latter is done with the \textsc{BELM} code.  At each radius, this code computes $q_{thin}$ and the associated spectrum for a sphere of radius $H(R)$. The radiation emitted by the corresponding disk annulus is then just the sum of all spheres filling in the same volume.  In this paper,  electrons and ions have been assumed to follow a thermal distribution with no pairs, no nonthermal particles and no external photon source. For more information on the code, we refer to \citet{Belmont08, Belmont09}.

Solving the equations at each radius with the \textsc{BELM} code would be too time consuming, however. We used a table of \textsc{BELM} simulations instead that provides the various cooling terms along with their associated spectral emission as a function of electron temperature $T_e$, Thomson optical thickness $\tau_T$ , and magnetic field strength (translated here into $\mu$). We then performed cubic interpolations on the logarithm of the table quantities to derive the desired values at any given radius.

%

\begin{figure}[h!] 
   \centering
   \includegraphics[width=\columnwidth]{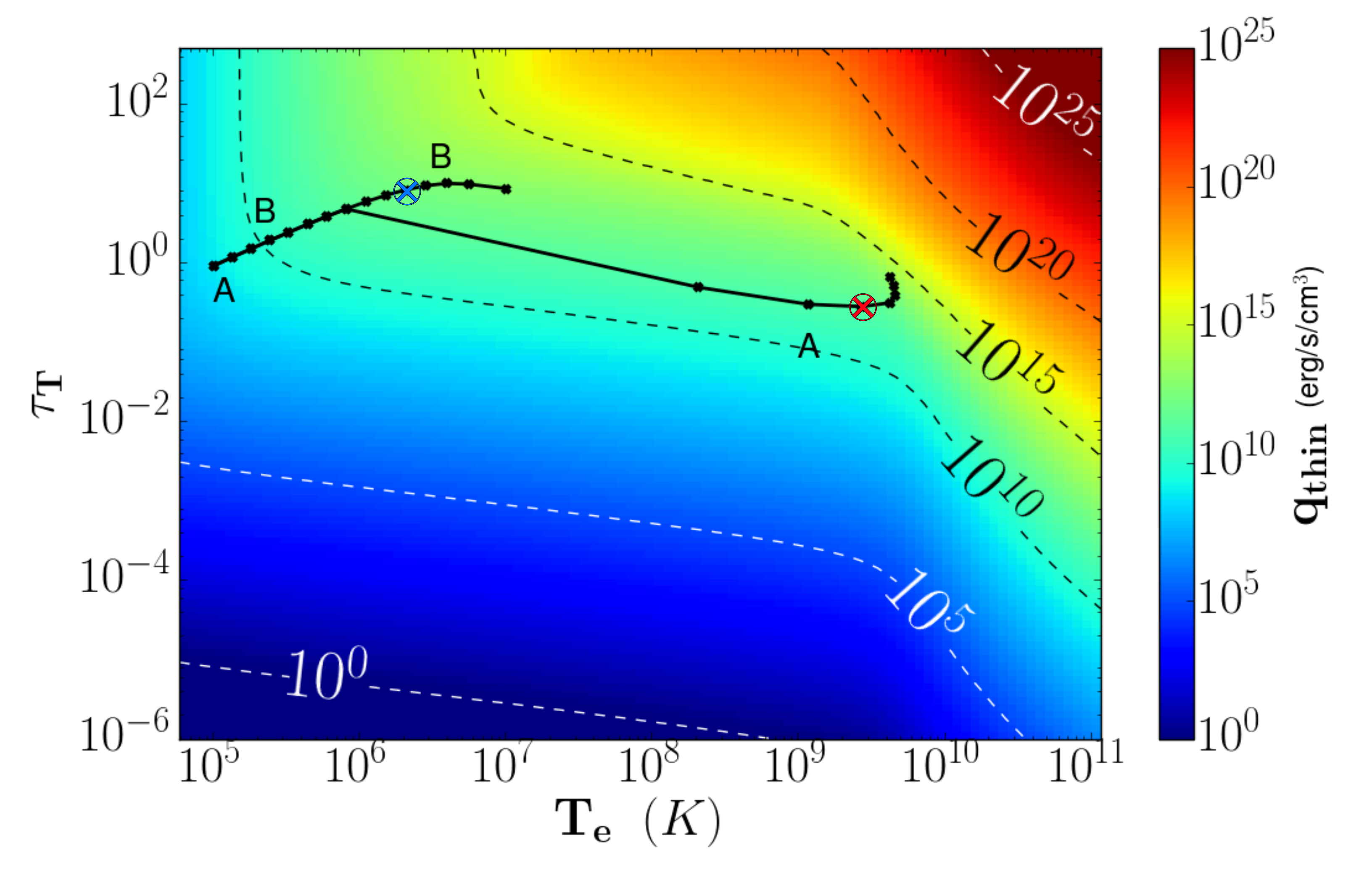}
      \caption{Map of interpolated optically thin radiative cooling term $q_{thin}$ (see section \ref{sec:BELM}) as a function of optical depth $\tau_T$ and electron central temperature $T_e$, computed for $\mu = 0.1$. The solid lines show the path in the $T_e-\tau_T$ plane followed by the two configurations A and B of the JED solution described in Sect.~\ref{sec:ExampleOfSolution}. Each dot along the lines corresponds to the radius where the disk thermal state has been computed, and the two cross markers correspond to the thick-disk (red) and thin-disk (blue) solution at $r=30$ (see text).}
         \label{fig:Interp}
\end{figure}

The \textsc{BELM} table was compiled using 4 values of $\mu$ ($0.01, 0.1, 0.5, 1$), 20 values of optical depth $\tau_{T}$ in the range $[10^{-6}, ~5~10^{2}],$ and 20 values of electron temperature $T_e$ within $[5~10^{4}, ~2~10^{11}]$~K. Figure~\ref{fig:Interp} displays an example of our interpolation of the total optically thin radiation term $q_{thin}$,  computed for $\mu = 0.1$ and the entire parameter space in $\tau_T$ and $T_e$. Although the table has only 20 values for each variable, the interpolated solution is extremely smooth so that using it produces no significant deviation from the exact solution. As a further illustration, we also plot (solid lines) the paths followed for the JED solutions described in Sect.~\ref{sec:ExampleOfSolution} from $r_{out} = 10^3$ to $r_{in} = 6$ (left to right) for the two configurations A and B. In addition, the two cross markers show the solution found at $r = 30$ (blue for thin, red for thick), used to detail the algorithm in the following paragraphs.\\

The advection term brings inwardly a fraction of the energy content of the outer disk regions. This term is very important, especially when the disk is hot (Sect.~2.3.3). In the literature, either global transonic calculations are made or some approximation on the radial derivatives is used in local equations \citep{Yuan01}. In order to consistently deal with this important effect, the resolution was performed outside-in, using third-order right-side derivatives approximations. As a consequence, computing the energy balance at a radius $R$ requires the knowledge of the thermal solution at the neighboring $R+dR$ radius as well as $R+2~dR$ and $R+3~dR$. This nonlocal effect needs to be included in our resolution scheme. We assumed the outermost parts of the disks (at $r_{out}$) to be in the cold SAD mode, in agreement with current theoretical expectations. The first advection term is then calculated using the outer SAD solution (through $\Delta_\alpha$, Eq.~\ref{eq:ADV_deltaalpha}). The code we developed solves this nonlinear system of three independant variables $(\varepsilon, ~T_e, ~T_i)$ using the following procedure. \\

The disk vertical equilibrium (eq.~\ref{eq:vert}) is solved first, using an iterative procedure providing the disk aspect ratio $\varepsilon$ for any given possible electron and ion temperatures $(T_e, ~T_i)$. Then, the ion and electron energy balance (eq.~\ref{eq:IONS} and eq.~\ref{eq:ELECTRONS}) are revised to provide a more straightforward resolution:
\begin{eqnarray}
     \left( 1-\delta \right) \cdot q_{turb} &=& q_{adv,i} + q_{ie} \label{eq:1} \\
      q_{turb} &=& q_{adv} + q_{rad} ~~~~~~~~~~ \text{} \label{eq:2}
,\end{eqnarray}
\noindent where eq.~(\ref{eq:1}) is still the ion energy balance, and eq.~(\ref{eq:2}) is now the total energy balance of the system $q_+=q_-$, obtained by adding eq.~(\ref{eq:IONS}) and eq.~(\ref{eq:ELECTRONS}).

\subsection{Multiple possibilities}

\begin{figure}[h!] 
   \centering
   \includegraphics[width=\columnwidth]{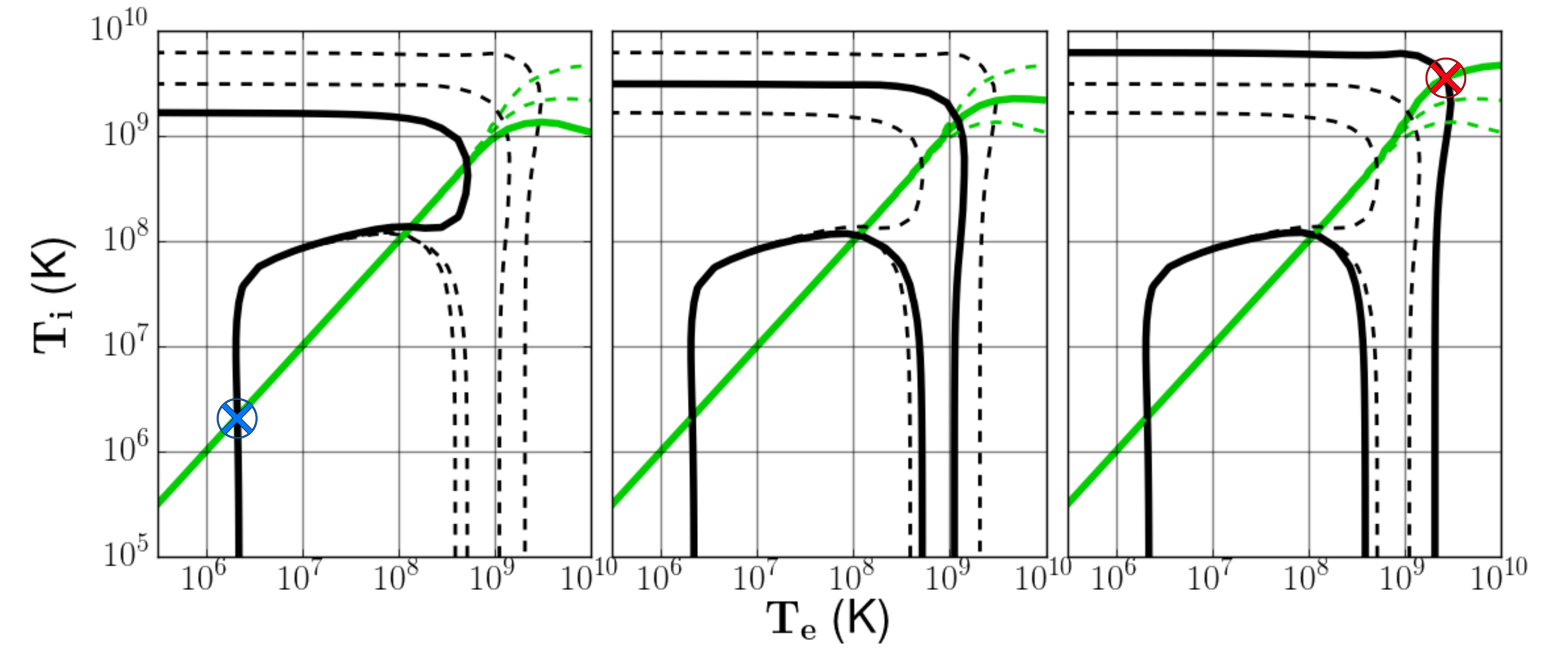}
      \caption{Zeros of eq.~(\ref{eq:1}) in black and eq.~(\ref{eq:2}) in green for each different dynamical solution at radius $r=30$ for the disk configuration from section~\ref{sec:ExampleOfSolution}. From left to right: thin disk, unstable disk, and thick disk. The zeros of the other dynamical solutions are displayed as dotted-lines for comparison.}
         \label{fig:AdvSol}
\end{figure}

At any given radius, one or three solutions are possible (see section \ref{sec:ExampleOfSolution}). The main complexity of computing a global disk thermal state configuration therefore was to handle possible optically thick-thin radial transitions. If there is only one possible solution, the inward computation is straightforward, the outer solution being used to compute the inner one, but when three solutions arise, the numerical scheme must keep track of these three solutions as long as they exist while going inward. This is simply because the advection term is now different for each branch. This implies following each of these solutions  separately and taking the different corresponding advection terms into account. In our resolution, the disk vertical equilibrium is solved for each $(T_e,~T_i)$ couple, narrowing our problem down to a 2D set of equations in the $(T_e,~T_i)$ plan. We show in Fig.~\ref{fig:AdvSol} the zeros of equations (\ref{eq:1}) and (\ref{eq:2}) as a function of $(T_e,~T_i)$ for each different outer solution of the disk: thin disk, unstable
disk, and thick disk. Any possible solution to the two equations at this given radius is thus given by an intersection between the green and black curves.

In this example, we see three different pictures. From left to right, the thin-cold branch, the unstable branch, and the thick-hot branch. From each of these branches, differences in the chosen solution branch provide different zeros for our two equations. Here, this gives rise to nine potential solutions: three different outer conditions, thin, unstable, and thick disk from left to right; and three different solutions, one for each different intersection of the green and black lines. However, only three solutions are physically consistent, each present solution associated with its progenitor: the thin (unstable, thick) disk solution associated to the thin (unstable, thick) outer conditions. The unstable solution is irrelevant, but the other two solutions are displayed in Fig.~\ref{fig:Interp} and \ref{fig:AdvSol} as the blue and red cross. \\

Finally, a fast, precise, self-consistent, and robust solving method was designed, providing not only all possible disk thermal states at each radius (temperature, geometry, etc.) but also the associated total spectrum. All of this on a large parameter space in disk accretion rate $\dot{m}_{in}$, JED parameters ($b$, $m_s$, $\mu,$ and $\xi$), outer SAD parameters ($\alpha_{\nu}$), and black hole binary parameters ($D$, $m$, $r_{in}$, and $r_{out}$). We note that for illustrative purposes, only $15$ to $20$ radial steps were used in the figures in this paper.  Our resolution was performed with a much higher precision, however, with $100$ radial steps.

\subsection{Discretization effect}
\label{sec:EffectofNdr}

Discretization is crucial in numerical simulations, and the number of steps is chosen as
\begin{eqnarray}
     N = 1+ \frac{\text{ln}(r_{out}/r_{in})}{\text{ln}(1 + dR / R)} \sim \frac{10}{dR / R} \label{eq:Ndr}
,\end{eqnarray}
\noindent with typical values of $r_{out}/r_{in} \sim 10^5$, assuming that $N \gg 1$ and $dR \ll R$. Obviously, mathematically speaking, better solutions are obtained as $N$ increases. However, in our model, each disk annulus at a given radius $R$ is replaced by a large number of spheres of radius $H(R)$ (see section \ref{sec:BELM}). The coherence between $H$, $dR$ and $R$ is fundamental, especially for optically thin solutions, and the best description should be obtained when $dR = H \simeq 0.1 R$ in JEDs, which leads to $N \sim 100$. \\

\begin{figure}[ht]
  \centering
  \includegraphics[width=\columnwidth]{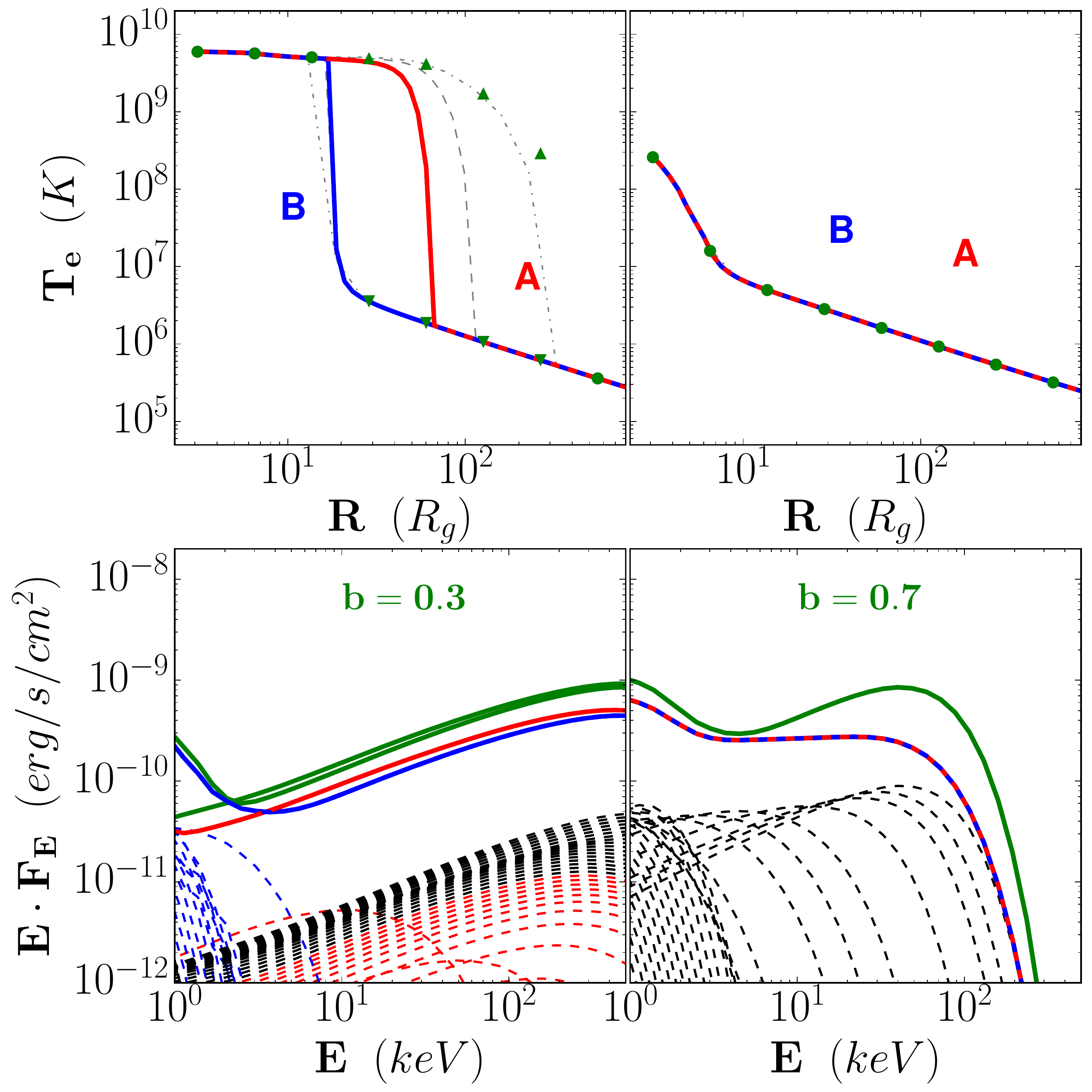}  
  \caption{Resolution effect on the radial thermal structure (top) and the global SED (bottom) for two values of the jet power fraction $b$. This figure is similar to Fig.~\ref{fig:EffectOfB}, which has been computed using $N=15$ and is reported here as green markers (top) and green solid lines (bottom). The results obtained for $N=40$ and $70$ are shown as dot-dashed and dashed gray lines in the top panel, respectively. The colors show solutions for $N=100$.}
  \label{fig:EffectOfB_npts}
\end{figure}

For the sake of illustrative purposes, we chose to display spectra with only $N=15$ to $20$ in Sections \ref{sec:JEDthermalstates} and \ref{sec:Params} (but calculations were made with $N=100$ in sections \ref{sec:Hysteresis} ans \ref{sec:bestparams}). Calculating the thermal equilibrium with such few steps imposes $dR / R \sim 1$. We are aware that in this case, the solutions obtained cannot be taken at face value (see Fig~\ref{fig:EffectOfB_npts}), but this is a pedagogic choice. More radial steps in figures \ref{fig:Exple} to \ref{fig:EffectOfXi} would have been confusing.

Such a low radial resolution biases the final shape of the global spectrum. This is illustrated in Fig~\ref{fig:EffectOfB_npts}, where calculations were made with various values of N. Two main differences are noticeable. First, the global architecture of the solutions (the $\text{three}$ thermal solutions at a given radius) is maintained, but with a strong influence on the range in radius and the temperature value for the optically thin/hot solution. Second, the shape of the global spectrum is strongly modified, it changes from $N=15$ to $N=100,$ as illustrated in the bottom row of Fig~\ref{fig:EffectOfB_npts}. Because the solutions are different at some radius, the spectral shape presents differences in structure where the possible solution produced is now optically thick and no longer optically thin, which has a small but non-negligible effect, as is shown with $b=0.3$. In addition, the more radial steps are solved, the less important each local spectrum becomes. Thus, obviously, the evolution of the disk thermal state is better described when more radial steps are made. This is shown for $b=0.7,$ where only one local spectrum dominated at high energy for $N=15$, while for $N=100,$ there are now multiple components and the bump at high energy is no longer present.

\section{Comparison with other models} 

This is the first time that a two-temperature plasma code is used for jet-emitting disks (JED). Our code was validated using previous one-temperature calculations in the JED case, as well as two-temperature calculations performed in the standard accretion disk (SAD) case. 

We recovered the main results obtained by \citet{Petrucci10} in the case of a JED mode. In their paper, the authors computed the thermal balance using a one-temperature approximation, a local prescription for the advection term, and followed \citet{Esin96} for the optically thin radiation. Their Fig.~2 compares quite successfully with our own Fig.~\ref{fig:ThickThinSlim} for parameter set (c). There are some discrepancies, however, which are due to our better treatment of radiation and advection and of the collisional coupling between ions and electrons. Differences can be seen either at large $\dot m_{in}$ and small radii (appearance of slim solutions for other parameter sets) and at low $\dot m_{in}$ at large distances (non-existence of optically thin solutions).   

We adapted our code in order to compute the thermal balance of a disk accreting in SAD mode. For such a disk, we have $\xi= 0$ (no mass loss so that $\dot m(r)= \dot m_{in}$), $b=0$ (all released power is dissipated within the disk), while the accretion sonic Mach number writes $m_s= \alpha_v \varepsilon$, where the turbulence strength is measured by the Shakura-Sunyaev $\alpha_v$ parameter. We note that within our prescription, $\alpha_v$ includes the derivative of the turbulent torque. As for the JED, we assume that the same fraction of turbulent energy is dissipated in the ions and electrons, that is, $\delta=0.5$, forbidding thereby ADAF solutions \citep[as also suggested in][]{Yuan14}. For the magnetic field strength, we used $\mu=0.1$ but verified that none of our results depends on its value. The reason is that all of our SAD solutions are optically thick (see below). 
The SAD turbulent heating term writes
\begin{equation}
q_{turb}  =   \frac{3 G M \dot{M}}{8 \pi H R^3} \left( 1 - \sqrt[]{R_{in}/R} \right)
,\end{equation}
making use of the no-torque condition imposed at $r_{in}$. All other terms appearing in the energy equation remain identical, the only difference with the JED is the value of $m_s$ and its dependence on the disk aspect ratio $\varepsilon=h/r$.

\begin{figure}[h!] 
   \centering
   \includegraphics[width=\columnwidth]{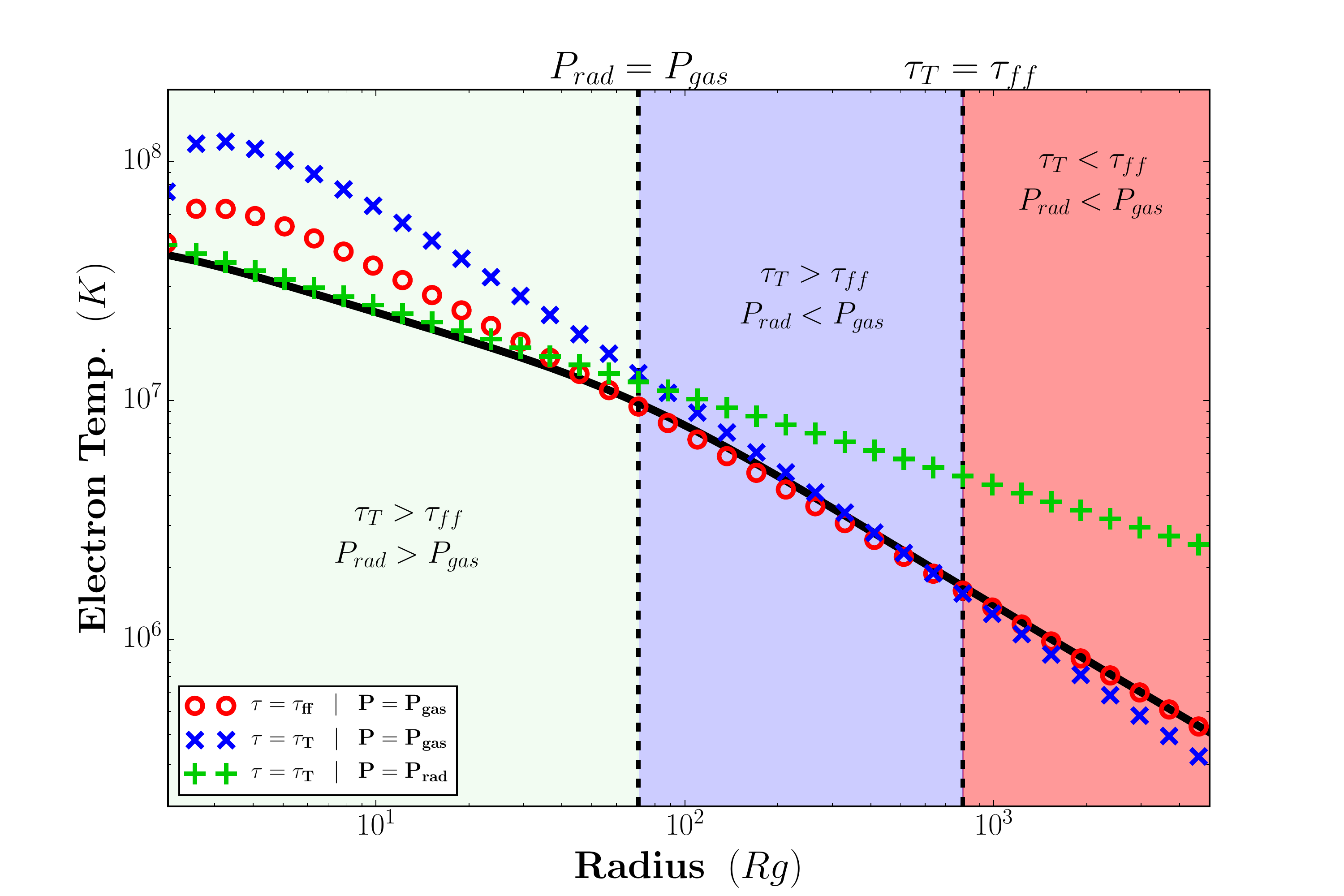}
      \caption{Comparison of the central disk temperature computed using our code (black solid line) with the analytical solutions, obtained for $\dot{m} = 1$, $\alpha_v = 0.1$ and $r_{in} = 2$. The three domains are shown as a background color: green ($P_{rad}, \tau_{tot}= \tau_T$), blue ($P_{gas}, \tau_{tot}= \tau_T$), and
red ($P_{gas}, \tau_{tot}= \tau_{ff}$). The analytical temperature profiles computed for each zone (symbols) are superimposed on the whole domain using the same color. The vertical dashed lines mark the transition radii between the zones.} 
         \label{fig:SAD}
\end{figure}

Figure~\ref{fig:SAD} shows the comparison of the disk central temperature computed using our two-temperature code with the analytical expressions derived in \citet{SS73} for the three innermost disk regions in the case of an optically thick, geometrically thin disk: (a) radiation pressure supported and Thomson opacity, (b) gas-pressure supported and Thomson opacity, and (c) gas-pressure supported and free-free opacity. The agreement is obviously excellent. The transition between each region (shown as color background) is smooth and occurs at the correct locations \citep[see their expressions in][]{Frank92}.

\begin{figure}[h!] 
   \centering
   \includegraphics[width=\columnwidth]{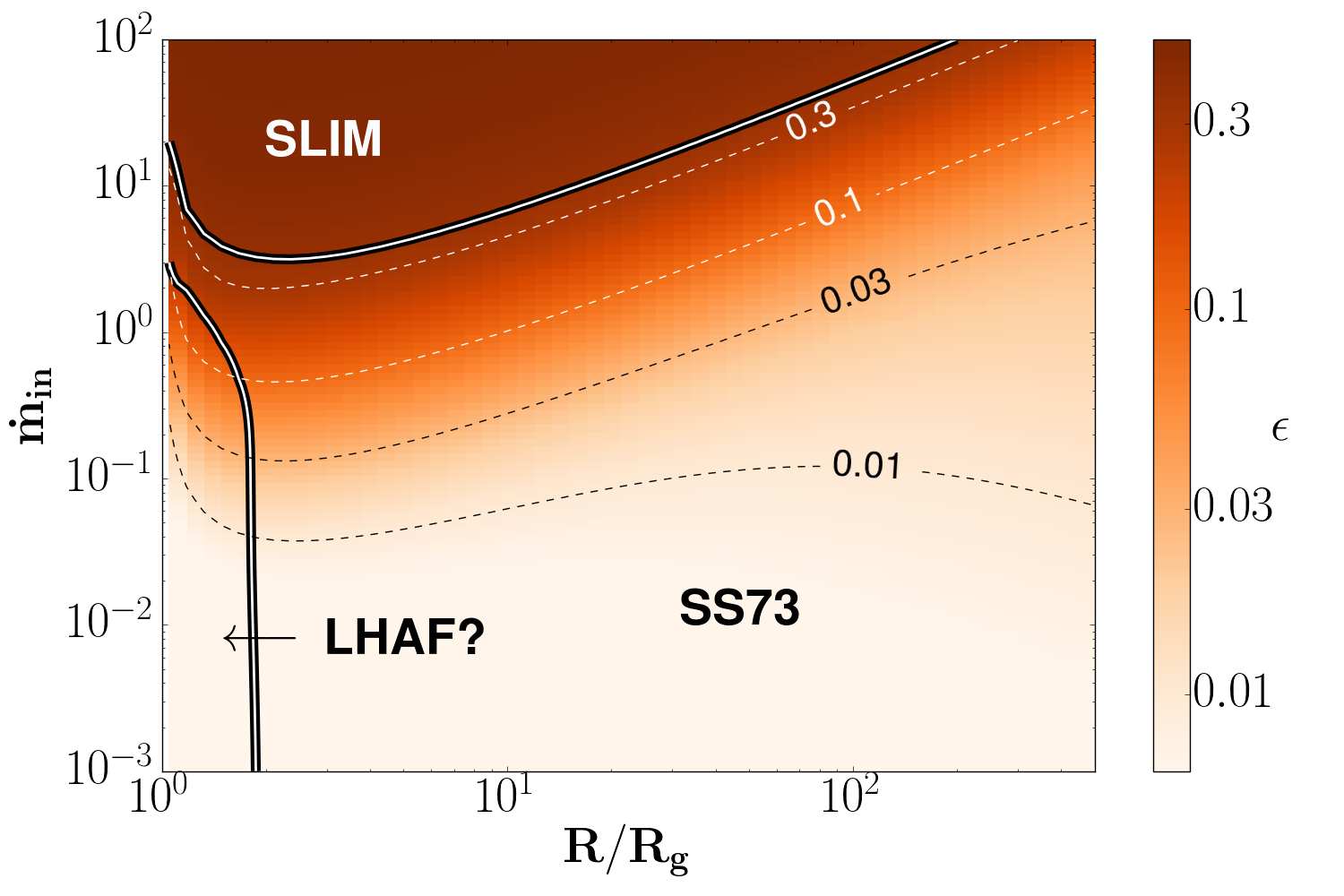}
      \caption{SAD ($\dot{m}-r$) solution plane for $\alpha_v=0.1$, $\mu=0.1$ and $r_{in}=1$. The background color and levels indicate the disk aspect ratio $\varepsilon=h/r$. The two bold lines are separatrices showing three different zones, each with its own spectral signature: the Shakura-Sunyaev (SS73) optically thick regime, the luminous hot accretion flow or LHAF (where $q_{adv} < 0$), and the SLIM disk where $q_{adv} > 0.5 q_{turb}$.}
         \label{fig:SADSlim}
\end{figure}

\begin{figure*}[ht] 
   \centering
   \includegraphics[width=\textwidth]{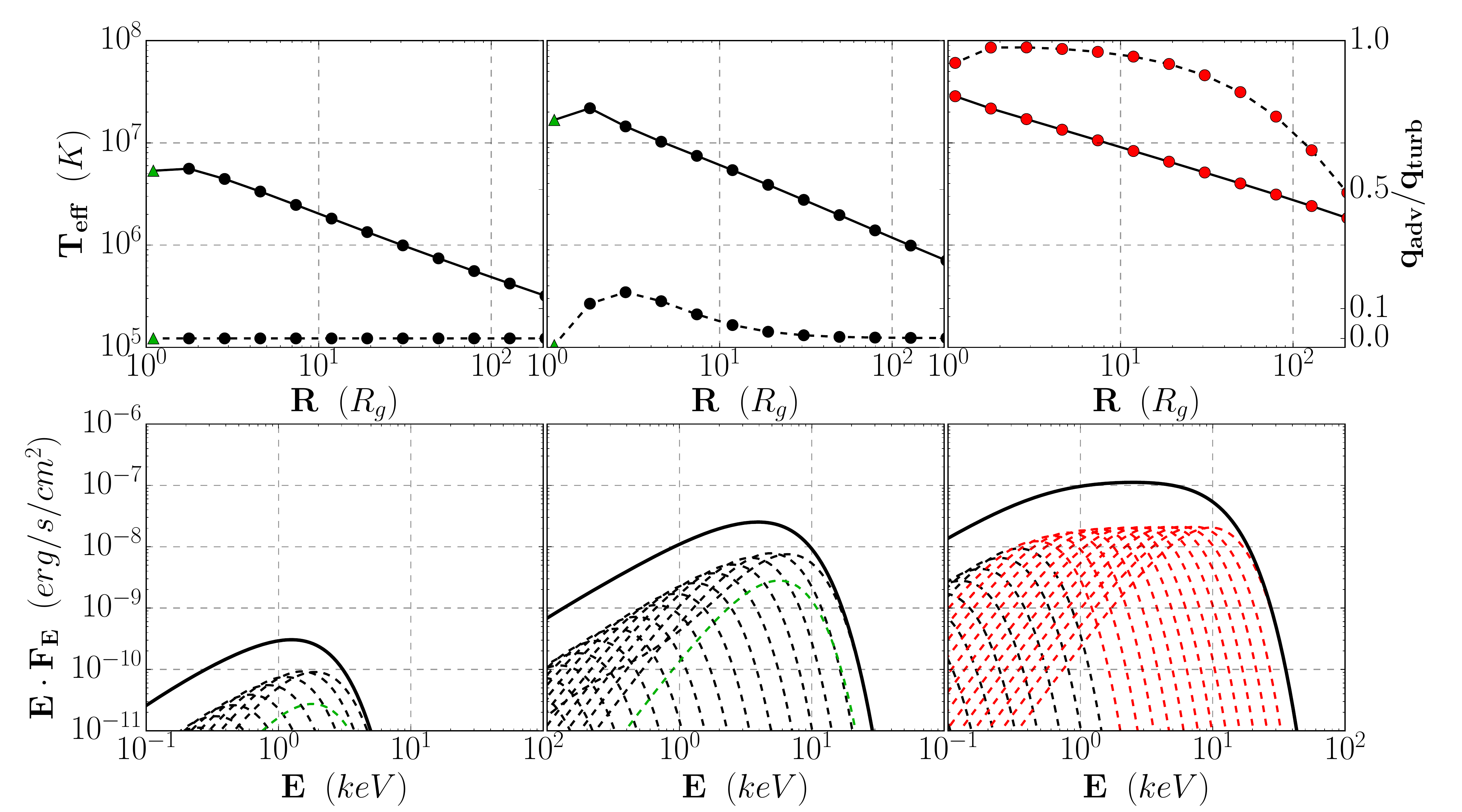}
      \caption{SAD configurations, settled from $r_{in}=1$ with $\alpha_v=0.1$, fed with an increasing $\dot m_{in}$ from left to right: $0.01, 1, 100$. The top panels show the disk effective temperature $T_{eff}$ (solid lines) and ratio $q_{adv}/q_{turb}$ (dashed lines). 
The bottom panels display the associated disk spectra. The dashed lines are the spectra emitted from each disk annulus whereas the solid black lines are the total integrated spectrum. A red color (in dots and spectra) describes annuli dominated by advection, whereas a green color (triangles and spectra) describes a negative advection term (see text).}   
         \label{fig:NEDSpectra}
\end{figure*}

In order to compare the results provided by our code with those shown in Fig.~2 in \citet{Chen95}, we computed global disk thermal structures with $\alpha_v=0.1$, $\mu=0.1$ and $r_{in}=1$ by varying the accretion rate $\dot{m}_{in}$ from $10^{-3}$ to $10^2$.  Our $(\dot{m}-r)$ parameter space is plotted in Fig.~\ref{fig:SADSlim} where three different zones can be distinguished. In order to better characterize them, we made a slice at a given $\dot m$ and displayed the radial distribution of their effective temperature (top panels) and associated spectrum (down panels) in Fig.~\ref{fig:NEDSpectra}. The results are described below.\\

\textbf{SS73:} The first zone is defined by the optically thick, geometrically thin ($\varepsilon \sim 0.01$) accretion disk. This SS73 domain corresponds to solutions with negligible advection, where turbulent heating is balanced by radiative diffusion and the global spectrum is the classical multicolor blackbody spectrum. The typical central temperature is $T_e=T_i \simeq 10^5-10^7$ K. An example is displayed in the left panel in Fig.~\ref{fig:NEDSpectra} with $\dot m_{in}=0.01$. It produces a very low luminosity spectrum with $L \simeq  6 \times 10^{-3} L_{Edd}$ and a maximum temperature of about $T_{eff,in} = 0.5$~ keV. \\

\textbf{LHAF:} The second domain is found mostly in the lower (low $\dot m_{in}$) part of the figure and always below $2R_g$. It corresponds to the LHAF solution \citep[luminous hot accretion flow,][]{Yuan01}, which is obtained here when the advection term $q_{adv}$ becomes negative (bold line). Instead of cooling the plasma, advection leads to local heating. This situation arises here because two elements are combined: (1) the zero-torque condition,  which enforces the disk turbulent heating to decrease abruptly, and (2) a low to moderate disk accretion rate that is not high enough to compensate for this. It is not clear to us whether such a solution would survive if a fully relativistic calculation were made, as implied by $R<2R_g$\footnote{We note that our calculations assume $m_s= \alpha_v \varepsilon$, that is, an accretion that
is always subsonic, in strong contrast with proper transonic LHAF solutions. However, the question of the influence of the outer boundary on the emergence of such solutions remains \citep[see, e.g.,][]{Yuan00, Artemova06}.}. An example is shown in the middle panel of Fig.~\ref{fig:NEDSpectra} with $\dot m_{in}=1$. It produces a spectrum quite similar to the previous one, however, with a higher luminosity  $L \simeq  0.6 \,  L_{Edd}$ and a temperature around $T_{eff,in} = 1.7$~ keV. \\

\textbf{Slim:} The last domain, in the upper (high $\dot m$) part of the figure, corresponds to the SLIM solutions where advection cooling becomes important \citet{Abramowicz80}. The bold separatrix line corresponds to $q_{adv}/q_{turb} = 50 \%$. This type of disk is optically thick $\tau_{tot} > 1$ and geometrically thick ($\varepsilon$ up to 0.4). Such slim solutions require high accretion rates and are thus dominated by radiation pressure. An example with  $\dot m_{in}=100$ is shown in the right panel of Fig.~\ref{fig:NEDSpectra}. Since the spectrum is dominated by the inner regions, the spectral shape is fundamentally different from that of a multiple blackbody. Despite the differences in our resolution methods, the computed spectrum displayed here is comparable to that shown Fig.~1 in \citet{Watarai01}. The luminosity obtained in this case is $L \simeq 4 \, L_{Edd}$.

To summarize, our two-temperature code recovers most of the previously published results for SAD and JED accretion modes, with an greater capacity to handle emitted spectra, however, since these are now consistently computed along the thermal balance of the disk. It is remarkable that according to its position in Fig.~\ref{fig:SADSlim}, the spectral signature of a SAD configuration will resemble one of the panels in Fig.~\ref{fig:NEDSpectra}. As expected, these signatures are unable to explain the complete ensemble of canonical states found in XrBs. While perfectly suited to explain the soft states, SAD configurations fail to explain the hard states. Even if the spectral signature of SLIM disks could be mistakenly interpreted as a power-law spectrum, a SAD configuration provides SLIM spectra only at very high luminosities, much higher than those currently obtained in XrBs. The vast majority of hard states is therefore beyond reach of SAD configurations. In our view, this is consistent with the underlying dynamics, as a SAD mode is also unable to produce jets and/or winds that are observed during the hard states.  Another dynamical accretion mode therefore needs to be invoked.   

\end{document}